\documentclass[]{aa}  

\usepackage{graphicx}
\usepackage{txfonts}
\usepackage{tikz}
\usepackage[colorlinks=true,linkcolor=blue,citecolor=blue,urlcolor=blue]{hyperref}
\begin{document}

   \title{Radio and high-energy variable emission from a multi-wavelength follow-up of the infant planet-hosting star V1298\,Tau}
   \author{M.~Damasso\inst{1}\fnmsep\thanks{The first two authors contributed equally to this work.}\fnmsep\thanks{Corresponding authors; mario.damasso@inaf.it, vigano@ice.csic.es}
          \and D.~Vigan\`o\inst{2,3}
          \and R.~Spinelli\inst{4}
          \and S.~Kaur\inst{5}
          \and A.~Petralia\inst{4}
          \and \'A.~S\'anchez-Monge\inst{2,3}
          \and A.~F.~Lanza\inst{6}  
          \and A.~Maggio\inst{4}    
          \and J.~M.~Girart \inst{2,3}
          \and F.~Borsa\inst{7}
          \and F.~Del Sordo\inst{8,6}
          \and G.~Ghirlanda\inst{7}
          }

\institute{INAF -- Osservatorio Astrofisico di Torino, Via Osservatorio 20, I-10025 Pino Torinese, Italy
\and Institut de Ci\`encies de I'Espai (ICE-CSIC), Campus UAB, Carrer de Can Magrans s/n, 08193 Cerdanyola del Vallès, Barcelona, Catalonia, Spain
\and Institut d’Estudis Espacials de Catalunya (IEEC), 08860 Castelldefels, Barcelona, Catalonia, Spain
\and INAF -- Osservatorio Astronomico di Palermo, Piazza del Parlamento 1, I-90134, Palermo, Italy
\and Departament d’Astronomia i Astrof\'{i}sica, Universitat de Val\`{e}ncia, C. Dr. Moliner 50, E-46100 Burjassot, Val\`{e}ncia, Spain
\and INAF -- Osservatorio Astrofisico di Catania, via Santa Sofia 78, I-95123 Catania, Italy
\and INAF -- Osservatorio Astronomico di Brera, via E. Bianchi 46, I-23807 Merate (LC), Italy
\and Scuola Normale Superiore, Piazza dei Cavalieri 7, 56126 Pisa, Italy
        }

   \date{Received; accepted}

 
  \abstract
   {V1298\,Tau is a very young and magnetically active K1V star which hosts a benchmark multi-planetary system to study planet formation and evolutionary history at the earliest stages. Because of the high interest around this system, it has been the target of a multi-wavelength follow-up so far.}
   {Between 2025 and 2026, we conducted a campaign to monitor the variability of V1298\,Tau in radio (band C), X-ray, and NUV bands, with the support of publicly available contemporary optical light curve, and to investigate multi-band correlations to link the emission with stellar activity. }
   {Radio observations were carried out with the Karl G. Jansky Very Large Array (JVLA) in band C (4.5--6.5 GHz) between the end of October 2025 and early February 2026. X-ray and NUV observations were carried out with the Swift satellite over 20 days, ending one month before the begin of the radio campaign. We analysed the temporal variability of peak (integrated) fluxes in the C-band and of X-ray (NUV) luminosities on various timescales, comparing it with the variability observed in the high-cadence optical light curve collected by the ASAS-SN survey.}
   {We observed both radio and high-energy time variable emission. In the radio, JVLA detected emission in 90\% of the visits, switching between high ($\geq$0.1 mJy/beam) and low flux states in matter of days or hours. The fluxes show moderate correlation with the optical light curve, suggesting modulation with the stellar rotation period. Except for one visit, dynamic spectral analysis does not reveal short-term (minutes) variability, nor is there any evidence of circularly polarised emissions. Overall, the radio observations can be explained as gyrosynchrotron and/or cyclotron emission from the corona, and linked to stellar magnetic activity. At high-energy, we detected significant variability in X-ray luminosity (within the range $\sim$1.2--2.5$\times$10$^{+30}$ erg/s), the decay phase of a stellar flare, and evidence of correlation with the optical flux. Our data confirm that V1298\,Tau follows the near-linear Güdel-Benz relation, with radio and X-ray luminosities consistent with those measured for magnetically active pre-main sequence stars.}
  {}
  
   
   \titlerunning{Radio and high-energy follow-up campaigns of V1298\,Tau}
   \authorrunning{Damasso et al.}
   \maketitle
%
\section{Introduction}

V1298\,Tau is an infant star 20$\pm$10 Myr old located at a distance $d=108.02\pm$0.23 pc which hosts a multi-planetary system, consisting of at least four transiting planets detected by the \textit{Kepler}/K2 space telescope \citep{2019AJ....158...79D,david2019ApJ...885L..12D}. It represents a benchmark among new born planetary systems to investigate the formation and evolution of planets within the first hundred Myr. In an attempt to improve the multi-wavelength characterisation of the host star, \cite{damasso25} (hereafter D25) reported the first detection of time variable radio emission from V1298\,Tau in L (1--2 GHz) and C (4.5--6.5 GHz) bands with the Karl G. Jansky Very Large Telescope (JVLA).
That study was motivated by the expectation of detectable non-thermal radio emission over a broad range of frequencies, triggered by intense stellar magnetic fields and dense winds which are predicted to characterise infant stars. Just to mention a few notable examples V\,830\,Tau, a 2 Myr-old T Tauri star at 130.37$\pm$0.28 pc, was detected as a radio-bright and variable source thanks to JVLA observations \citep[detected emission larger than 1 mJy in band C;][]{Bower_2016,osten25}, and emission was also recorded from the multi-planetary system AU Mic (age $\sim$22 Myr old), monitored during an intensive radio campaign with the Australia Telescope Compact Array between 1.1 and 3.1\,GHz \citep{aumic24}. In general, the increasing survey coverage and sensitivity allow to radio detect many young, nearby stars \citep{Launhardt_2022,driessen24}. The radio-monitoring conducted by D25 on V1298\,Tau was intended as a pilot targeted follow-up to increase the radio sensitivity with respect to survey-like campaigns, to contribute to a first characterisation of the plasma environment where the planets are embedded, and to probe the strength of the coronal magnetic field and plasma density. With a similar motivation, \cite{Maggio2023} monitored the far-UV and X-ray high-energy emission from this system with the Hubble Space Telescope/Cosmic Origins Spectrograph and XMM-Newton, investigating the time evolution of the irradiation in the UV band, and its effect on the evaporation of the atmosphere of the infant exoplanets at the very early stages.

As a continuation of a characterisation study of the radiation environment and stellar activity variability in the V1298\,Tau system, we present results from a contemporaneous multi-wavelength follow-up conducted at radio frequencies with the JVLA (with more observations and a better sampling than in D25), and in X-ray and near-ultraviolet (NUV) with the Neil Gehrels Swift Observatory, supported by the publicly available archival optical light curve ($g$-band) collected by the All-Sky Automated Survey for Supernovae (ASAS-SN).

\section{Observation set-up and data processing} \label{sec:dataset}

\subsection{Radio follow-up} 

The target was observed with JVLA in band C ($4.5-6.5$ GHz) during 20 slots, from 30 October 2025 to 3 February 2026, with a dense sampling of 14 observations concentrated in 16 days during January. Each visit was about 1\,hour long, consisting of $\sim$8 minutes per each of the four target scans, separated by $\sim 2.5$ minutes phase calibrator scans, for a total of $\sim$35 minutes on-source. 
Full-Stokes imaging observations were conducted in standard continuum mode using the JVLA B-configuration, except for the last four, which were taken in the B$\rightarrow$BnA, B$\rightarrow$BnA, BnA, and BnA configurations, respectively. The spectral set-up consisted of 16 spectral windows with 64 channels of 2\,MHz spectral resolution, for a bandwidth of 128\,MHz, and a total bandwidth of 2\,GHz. We observed 3C147 as absolute flux and bandpass calibrator while J0403$+$2600 was used for gain and phase calibration. Data were flagged and calibrated using the JVLA calibration pipeline \citep{vla-pipeline}, after which we performed standard manual sanity checks of data quality mostly in the phase calibrator scans. The fraction of flagged data is between 19$\%$ and 30$\%$, depending on the observation, with excluded data having in common the entire spectral window centred around 5.8 GHz. We measured peak and integrated fluxes for each image using the Gaussian fitting procedure in Cube Analysis and Rendering Tool for Astronomy \citep[{\tt CARTA}; ][]{carta_1,carta_2}. Operative antennas were always between 25 and 27. Typical beam sizes are around $(1-1.3)''\times(1.4)''$, with the last (B$\rightarrow$)BnA observations much more elongated.

\subsection{X-ray and NUV follow-up}\label{sec:X}
V1298\,Tau was observed with the Neil Gehrels Swift Observatory \citep{Gehrels2004} in the X-ray and near-ultraviolet (NUV) bands. The dataset comprises a dedicated campaign of nine visits conducted in September 2025. The exposure times varied between $\sim$1.3 and $\sim$2.9 ks per visit.

For the X-ray analysis, we retrieved the science-ready images, light curves and spectra for each observation directly from the UK Swift Science Data Centre online facility \citep{Evans2007, Evans2009}. The X-Ray Telescope (XRT, \citealt{Burrows2005}) spectral analysis was performed in the 0.3--10 keV energy band using the \texttt{xspec} package \citep[version 12.15.0,][]{Arnaud1996}. We fitted the spectra using an absorbed, optically thin thermal plasma model (\texttt{tbabs $\times$ apec}; \citealt{Smith2001}).

The abundance table of \cite{wilms2000} is adopted to compute the absorption. During the spectral fitting procedure, the interstellar hydrogen column density ($N_{\rm H}$) and the stellar coronal metallicity were fixed to $1.6 \times10^{20}$\,cm$^{-2}$ and 0.2 $Z_{\odot}$, respectively, following \cite{Maggio2023}. 

The thermal emission is described by the plasma temperature ($kT$) and the emission measure ($EM$). The latter is calculated from the \texttt{apec} normalization parameter $N$ (which has dimensions [length]$^{-5}$) via the relation $EM = 10^{14} \, N \, 4\pi d^{2}$, where $d$ is the distance to the star, and $EM = \int n_e n_H dV$, where $n_e$ and $n_{H}$ are the electron and hydrogen densities, respectively, and $dV$ is the element of the X-ray emitting coronal volume.

We tested both single-temperature (1T) and two-temperature (2T) plasma models. To statistically evaluate the need for an additional thermal component, we computed the Akaike Information Criterion (AIC, \citealt{Akaike1974}) for both models. A 2T model was favoured over a 1T model only if it yielded an improvement of $\Delta{\rm AIC} \ge 10$. Based on this threshold, all the observed X-ray spectra were adequately described by a 1T model. This is expected, given the relatively low number of counts which prevent to characterize the spectrum with higher precision. The source unabsorbed flux is obtained by integrating the best-fitting {\tt apec} model over the 0.3-10 keV observation band. Best-fit parameters and their error estimates are computed via Markov Chain Monte Carlo within {\tt xspec}.

Simultaneously, NUV photometry was acquired using the Swift UV/Optical Telescope (UVOT, \citealt{Roming2005}) in the UVM2 filter (central wavelength $\lambda_c$=2246$\,\AA$). The UVOT data were processed and analysed using the public \texttt{heasoft} package and the most recent version of the calibration database (CALDB). Source photometry was performed using the \texttt{uvotsource} tool within a circular extraction region of 5 arcsec in radius, centered on the celestial coordinates of V1298\,Tau. The background level was evaluated from a source-free circular region of 20 arcsec in radius located in close proximity to the target. Finally, to derive the NUV luminosities ($L_{\rm NUV}$), assuming a flat spectrum in erg cm$^{-2}$ s$^{-1}$ \AA$^{-1}$, the measured count rates were multiplied by the appropriate flux density conversion factor ($7.47 \times 10^{-15}$ erg cm$^{-2}$ s$^{-1}$ \AA$^{-1}$ counts$^{-1}$, \citealt{Brown2016}) and by the filter full width at half maximum (FWHM = 498 \AA) to obtain the integrated energy flux.

\subsection{ASAS-SN optical light curve} \label{sec:asas}
We used the optical light curve (\textit{g}-band) of V1298\,Tau collected by the ASAS-SN\footnote{The data are publicly available at \url{http://asas-sn.ifa.hawaii.edu/skypatrol/}.} \citep{2014ApJ...788...48S,2023arXiv230403791H} during the time span covered by Swift and JVLA observations, to monitor the photospheric variability and investigate any correlation with radio and high-energy fluxes, similarly to what has been done by D25 with measurements of the first radio campaign. \\

\begin{figure}
    \centering
  \includegraphics[width=0.95\linewidth]{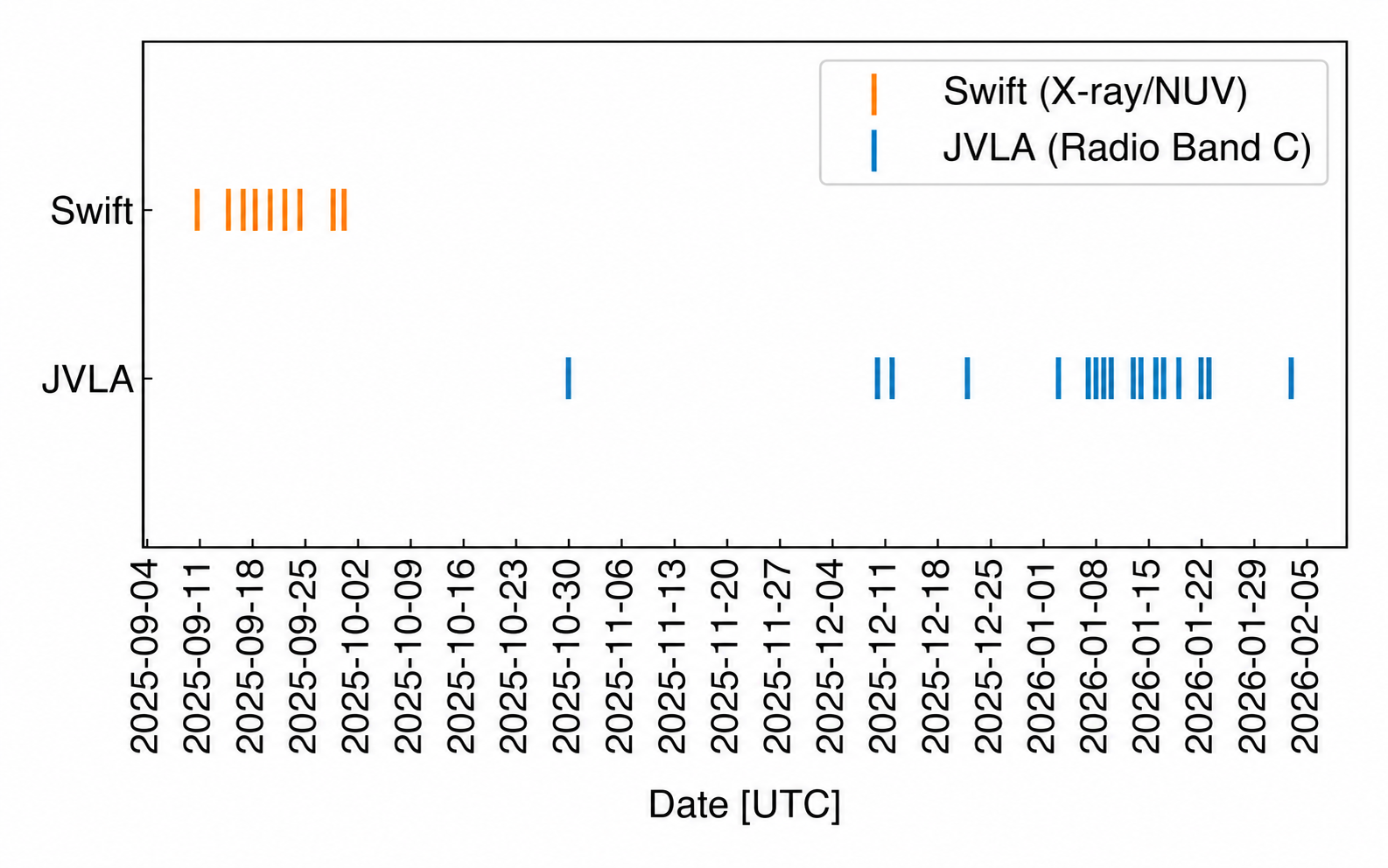}
    \caption{Timeline of the observing campaigns on V1298\,Tau conducted between September 2025 and February 2026 with Swift and JVLA. Orange vertical ticks represent the Swift X-ray and NUV observations, while blue ticks indicate JVLA radio observations in Band C. Data from ASAS-SN are available along the full timespan.}
    \label{fig:summarylog}
\end{figure}

Figure \ref{fig:summarylog} summarises the epochs of the JVLA and Swift observations of V1298\,Tau. The archival ASAS-SN light curve that we consider in our study spans the total time range covered by the two campaigns. On average, three observations every $\sim$0.5 d were collected by the ASAS-SN survey, with a median temporal separation of $\sim$1.5 d between consecutive observations after binning (bins of 0.5 d). X-rays, NUV and radio fluxes were converted into luminosities assuming isotropic emission and adopting V1298\,Tau's distance. These luminosities represent the average emitted power over the exposure time interval of each observation.

\begin{table*}
\centering
\tiny
\caption{JVLA observations of V1298\,Tau in Band C (4.5--6.5 GHz). }
\label{tab:radioresults}
\begin{tabular}{c c c c c c c c c}        
\hline\hline                 
Observation & Obs. start & MJD-2400000   & $\sigma$ & Peak flux\tablefootmark{a,b}  & Integrated flux\tablefootmark{b} \\
 & [yyyy-mm-dd hh:mm:ss] & [day] & [$\mu$Jy/beam]   & [$\mu$Jy/beam] &  [$\mu$Jy] \\
\hline
1 & 2025-10-30 05:02:00 & 60978.208 & 6 & $34 \pm 6$ & $35 \pm 10$ \\
2 & 2025-12-10 01:17:35 & 61019.054 & 6 & <18 & - \\
3 & 2025-12-12 07:20:58 & 61021.306 & 6 & <18 & - \\
4 & 2025-12-22 07:03:30 & 61031.294 & 6 & $47\pm 5$ & $27 \pm 6$ \\
5 & 2026-01-03 06:44:22 & 61043.281 & 6 & $150\pm 6$ & $138 \pm 10$ \\
6 & 2026-01-07 00:08:05 & 61047.006 & 6 & $127\pm 6$ & $153 \pm 12$ \\
7 & 2026-01-08 01:45:23 & 61048.073 & 9 & $360\pm 10$ & $365 \pm 18$ \\
8 & 2026-01-09 02:44:53 & 61049.114 & 21 & $181\pm 21$ & $149 \pm 33$ \\
9 & 2026-01-10 01:59:02 & 61050.083 & 6 & $122\pm 6$ & $120 \pm 11$\\
10 & 2026-01-10 23:58:30 & 61050.999 & 6 & $77\pm 7$ & $71\pm 11$ \\
11 & 2026-01-13 01:30:02 & 61053.063 & 6 & $100\pm 6$ & $90\pm 9$ \\
12 & 2026-01-14 04:56:12 & 61054.206 & 6 & $63\pm 6$ & $54\pm 8$ \\
13 & 2026-01-16 22:40:13 & 61056.945 & 6 & $25\pm5^c$ & $16\pm6$ \\
14 & 2026-01-17 05:53:29 & 61057.245 & 6 & $168\pm 6$ & $164\pm 9$ \\
15 & 2026-01-17 23:00:32 & 61057.959 & 6 & $25\pm 4^c$ & $12\pm5$ \\
16 & 2026-01-19 22:46:36 & 61059.949 & 6 & $80\pm 7$ & $95\pm 14$ \\
17 & 2026-01-22 00:25:18 & 61062.017 & 6 & $54\pm 6$ & $50\pm 10$ \\
18 & 2026-01-22 04:10:36 & 61062.174 & 6 & $27\pm 6$ & $44\pm 15$ \\
19 & 2026-01-23 00:17:00 & 61063.010 & 6 & $35\pm 4$ & $38\pm 9$ \\
20 & 2026-02-03 21:57:05 & 61074.915 & 6 & $197\pm 5$ & $209\pm11$ \\ \hline
All combined & - & - & 1.5 & $81\pm2$ & $76\pm 3$\\
\hline
\hline
\end{tabular}
\tablefoot{\tiny
\tablefoottext{a}{Upper limits are given as 3$\sigma$ confidence levels. The rms of Stokes I and V are always very similar.}
\tablefoottext{b}{Peak and integrated fluxes are extracted from the Stokes $I$ image plane, fitting the source emission with a 2D Gaussian.}
\tablefoottext{c}{These two detections are at the limit of the reliability, with a S/N $\sim$4 in the peak flux, but a smaller value of the integrated flux, probably due to the difficulty in the fitting of such a weak signal.}
}
\end{table*}

\section{Results} \label{sec:results}
\subsection{Radio C-band (4.5–6.5 GHz)} \label{sec:radioresults}
\subsubsection{Time variability}

\begin{figure}
    \centering
    \includegraphics[trim=0cm 0cm 0cm 1cm, clip=true, width=0.45\textwidth]{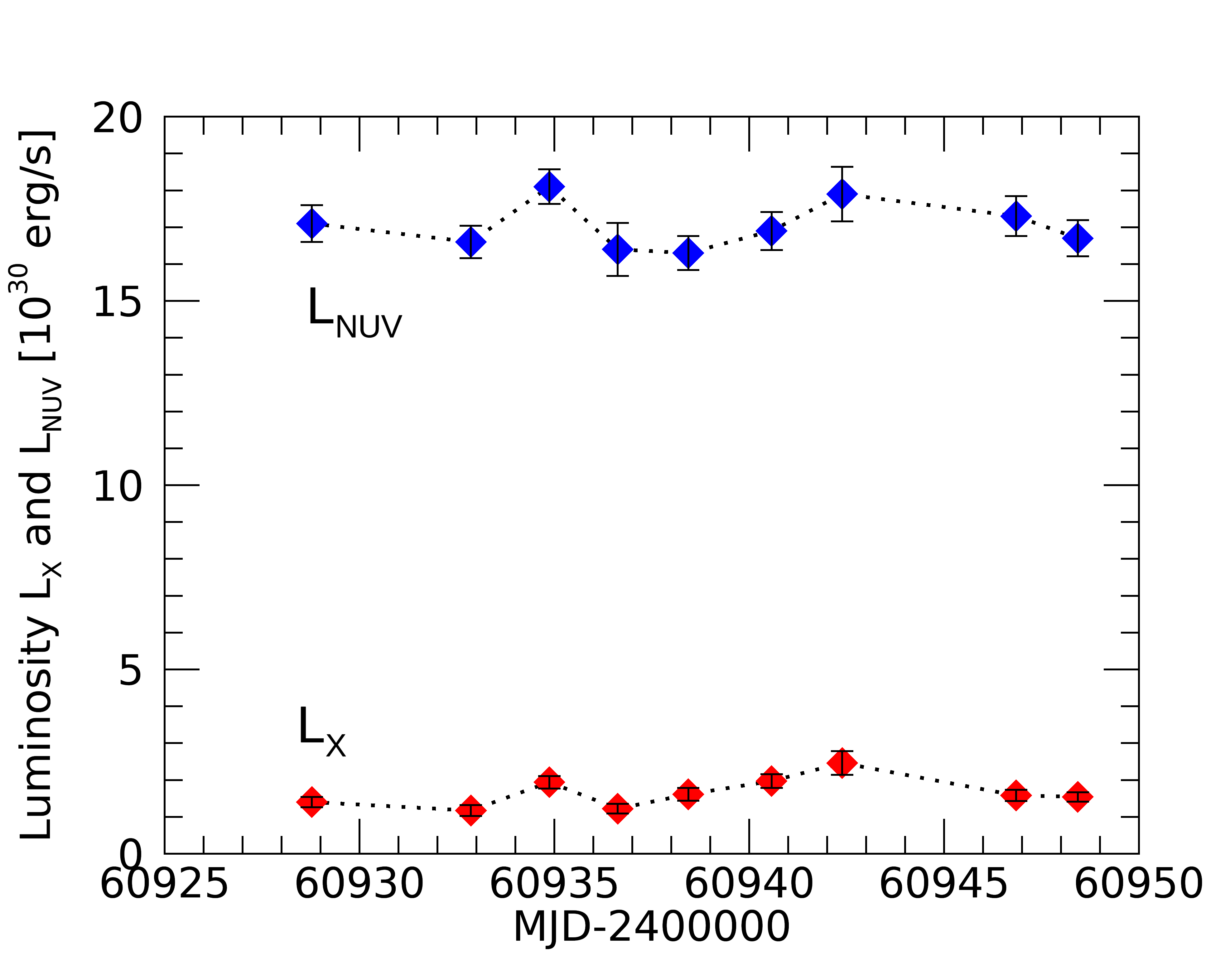}\\
    \includegraphics[trim=0cm 0cm 0cm 1cm, clip=true, width=0.45\textwidth]{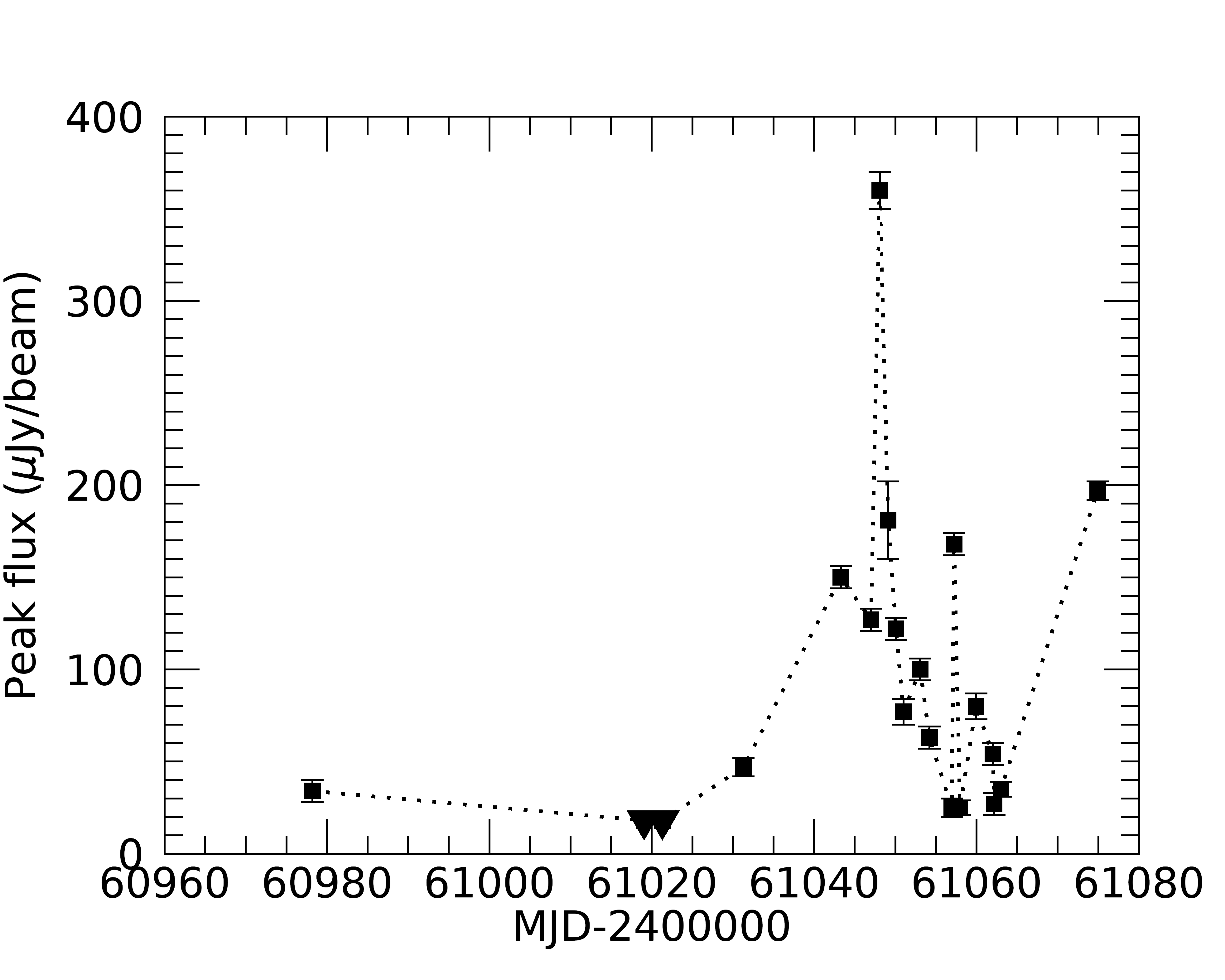}
    \caption{Time series of Swift (first panel; red and blue diamond for X-ray and NUV, respectively) and JVLA (second panel; radio band C) observations. Upper limits of radio fluxes are indicated with black triangles.}
    \label{fig:jvlatimeseries}
\end{figure}

The results of the JVLA follow-up are summarised in Table \ref{tab:radioresults}, where we show the background noise ($\sim$6 $\mu$Jy/beam, except for two epochs with $\sim$9 and 21 $\mu$Jy/beam), the peak and integrated fluxes. The peak flux time series is shown in the second panel of Fig. \ref{fig:jvlatimeseries}. We detect clear emission ($>$3$\sigma$ in both peak and integrated flux) in 16 out of 20 epochs. In two epochs (observations 13 and 15), there are $\sim4\sigma$ peak flux detections. Finally, only in observations 2 and 3, no hints for emission is seen (we represent these data with their 3$\sigma=18$\,$\mu$Jy/beam upper limits). Peak and integrated fluxes are always compatible with each other,\footnote{In dimmest epochs (peak fluxes $<50$\,$\mu$Jy/beam), the reconstructed integrated flux values suffer from higher fitting uncertainties due to the higher noise, so that their deviations from the peak flux values (up to a factor 2 higher or lower, see observations 4, 13, 15, 18) are not reliable.} indicating a spatially unresolved source, which is, not surprisingly, a loose constraint (emitting region $\lesssim 100$ au) compatible with any stellar-related emission, e.g. from stellar spots or magnetospheric belts.

When combining all epochs (last row of Table \ref{tab:radioresults}), the average emission is $\sim 80$\,$\mu$Jy/beam. However, there is a flux variability of up to one order of magnitude, if one compares the brightest epoch (360 $\mu$Jy/beam on 8 January 2026, almost twice the value of the second brightest that was recorded 23 hours later, observations 7 and 8) and the weak/non-detections. As a sanity check for such variations, we made sure that the few similarly bright background sources in the field of view (peak fluxes $\sim$0.5-1 mJy/beam) show much smaller ($\lesssim$10\%, and often negligible) variations among the different epochs (being more pronounced for the ones with the most elongated beams). Seven observations show peak (or integrated) fluxes above 100 $\mu$Jy/beam, five of which are consecutive and happening in one week, between 3 and 10 January (observations 5-9). The distribution of $\sim 1/3$ bright and $\sim 2/3$ dim fluxes is in line with our previous observations in the same band (D25), when two out of six observations showed fluxes $\gtrsim 100$ $\mu$Jy/beam, and the remaining had weaker values or non-detections, possibly representative of a quiescent state $\lesssim$50~$\mu$Jy. In this follow-up campaign, we also detected intermediate values (peak fluxes $\sim$70-95~$\mu$Jy/beam) in three cases, which partially fill the gap between the apparent bright-dim state dichotomy.

The high cadence of the observations collected in January allow us to appreciate flux variability on a relatively short time scale. With several consecutive days of monitoring, we notice significant variations on a day-by-day basis between 7 and 11 January, or the factor two difference in peak flux between the three observations taken within 24 hours on 22-23 January. The most evident change occurs within 24 hours between the three observations taken on 16 and 17 January (observations 13--15), during which the peak flux increased from $\sim$25 $\mu$Jy/beam to $\sim$170 $\mu$Jy/beam in just seven hours, and after 16 hours it returned to the previous barely detectable level.

We calculated the generalised Lomb-Scargle periodogram \citep[GLS][]{gls2009A&A...496..577Z} of the radio data, which shows the main peak at $\sim$28 d (comparable to timescale including the bulk of the observations) characterised by a high false alarm probability (FAP=21\%) assessed through a bootstrap analysis with 10\,000 permutations. The lack of significant signal is anyway not surprising, giving the small coverage fraction and restricted number of observations. About one order of magnitude more observations would be needed to have firmer conclusions, as \cite{aumic24} did for AU Mic.

\begin{figure}
    \centering
    \includegraphics[width=.9\linewidth]{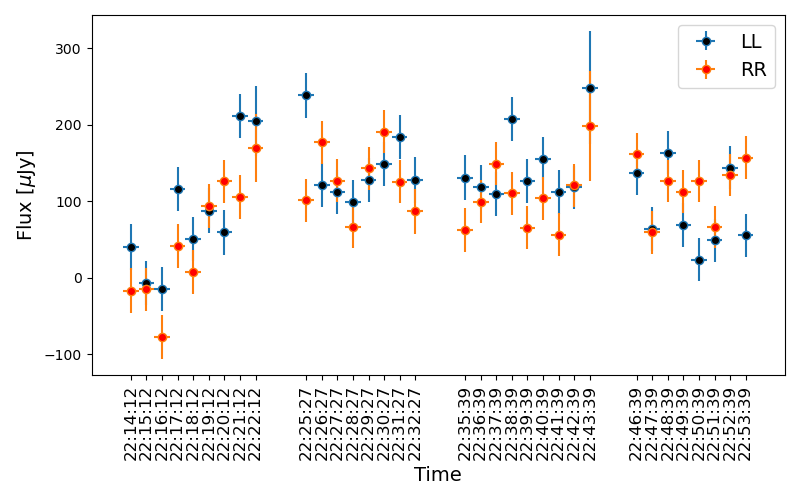}
    \caption{Variations of RR and LL radio flux within observation 20, obtained from the visibility plane. The statistically significant short-term variation is clearly visible also in the scan-by-scan Stokes I image comparison (not shown here).}
    \label{fig:radio_variations}
\end{figure}

We have also checked for short-term variability within each visit, analysing the clean images obtained for each of the four 8-minutes-long target scans in each observation. Only in one case (observation 20, the last one), we find a statistically significant ($>3\sigma$) scan-by-scan variation, with the two central scans being brighter than the first one ($\sim$50-70\% in flux), and the last one showing a mild decrease. As a confirmation of what is seen from the image plane, Fig. \ref{fig:radio_variations} shows the 60 seconds-binned light curve, as obtained from the visibilities, using the task {\tt visstat} and after subtracting, via the {\tt uvsub} task, the closest bright source NVSS
J040522+200958 (see App. \ref{app:maps}). A clear rise in both RR and LL components\footnote{RR and LL are the right-hand and left-hand circular polarization outputs provided by the correlator, e.g.: \url{https://science.nrao.edu/facilities/vla/docs/manuals/propvla/FAQs}. Their sum is proportional to the total intensity (Stokes $I$).} is seen at the beginning of the observation, with an increase from a dim state, compatible with the lowest or non-detected fluxes (few tens of $\mu$Jy/beam), to a bright state ($\sim 200$\,$\mu$Jy/beam). Time binning shorter than the 60 seconds does not show evidence of further variations, thus excluding very short, intense bursts. In all other observations, there are no or weak hints ($\sim 2\sigma$) for similar time variations, similarly to what shown in Fig. 3 of D25. Therefore, fluxes obtained from the $\sim$35 minutes integrated images are in general unlikely to be the product of isolated bursts with duration much shorter than the observations.

All together, the information we collected from these and past (D25) observations (26 epochs in total, in Band C) seems to indicate that the source jumps between different levels of emission, which are maintained steadily for hours at least, if not several days (observations 5-9 mentioned above). We don't see clear shorter ($\sim $ minutes) intense bursts. However, if the flux rise seen at the beginning of the last observation is representative of a state change, it could happen as fast as in a few minutes. 

Concerning variability over a longer time scale, the time series shown in Fig. \ref{fig:jvlatimeseries}, starting from observation 6, could be indicative of a long-lasting radio flare, as those detected in other young stellar systems \citep{osten09}, in particular V\,830\,Tau, for which a more luminous flare-like signal was recorded by JVLA in band C along $\sim$6 days \citep[][see Fig. 2 and 4 therein]{osten25}.

\subsubsection{Spectral analysis and polarisation}

\begin{figure}
    \centering
    \includegraphics[width=0.98\linewidth]{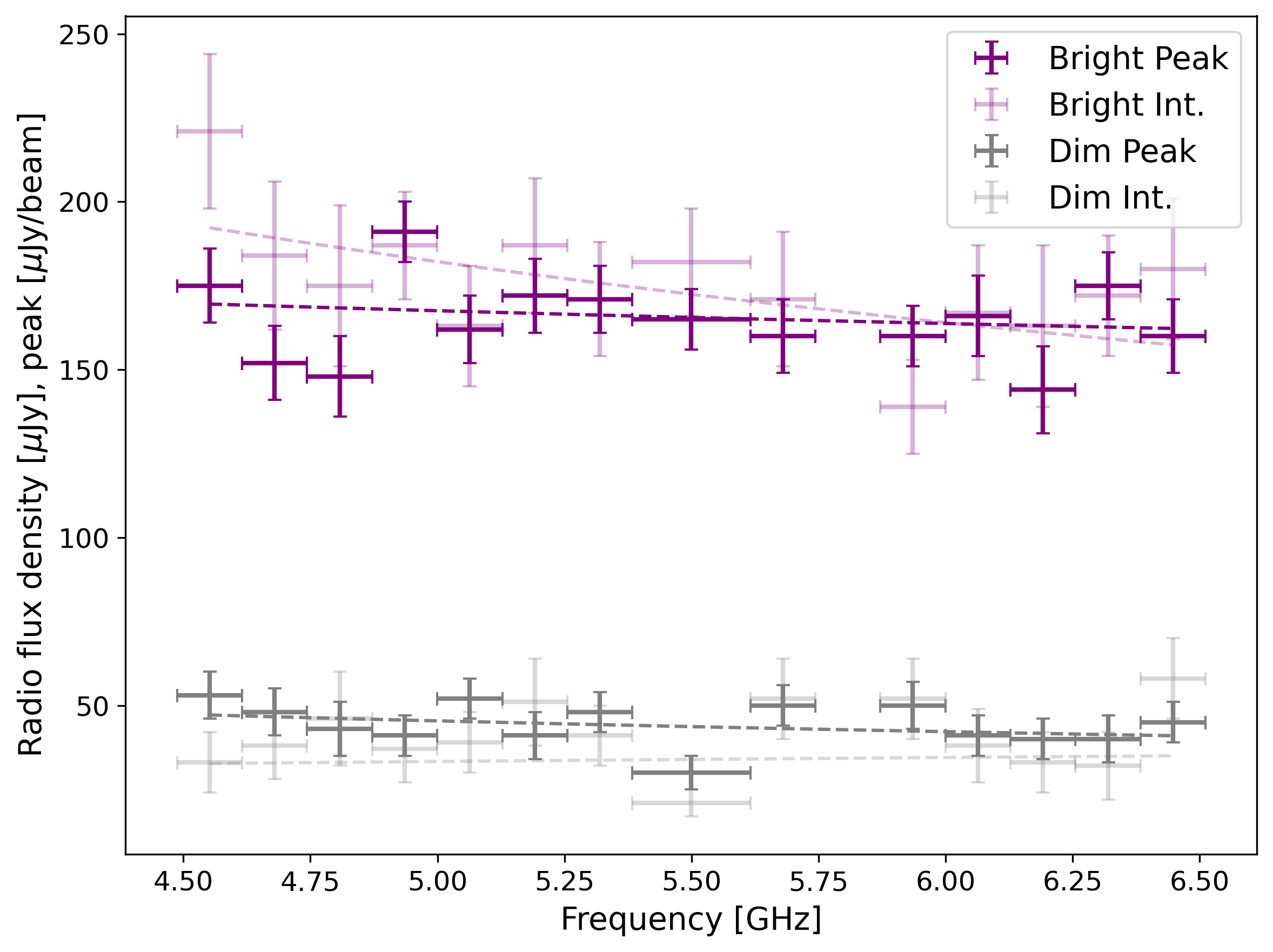}
    \caption{Spectra of the seven brightest epochs combined (integrated fluxes $>$ 100 $\mu$Jy, observations 5-9, 14, and 20; purple), and the remaining thirteen dim epochs (grey), considering the peak (dark) or integrated (light) Stokes I fluxes. The dashed lines identifies the corresponding best-fitting power laws. The gap around 5.8 GHz reflects the entire spectral window 10 flagged at all epochs, while the bin around 5.5 GHz has been taken wider than the rest (which are set at the spectral window bandwidth, 128 MHz), in order to compensate the many flagged channels in spectral windows 7 and 8.}
    \label{fig:radiospectra}
\end{figure}

In all cases, the radio emission looks broad across the 4.5-6.5 GHz band. Figure \ref{fig:radiospectra} shows the average spectra we obtain when we combine the seven brightest epochs (peak fluxes $>$100 $\mu$Jy/beam, purple), or the remaining thirteen dim epochs (grey). Spectra are produced from the clean cube image, with a user-defined mask and a threshold of $\sim 3\sigma\sim$ 30 and 20 $\mu$Jy/beam (bright and dim combined epochs, respectively). The best-fit power law seems to be slightly negative, but consistent with being flat, for both bright and dim epochs (spectral index for peak flux fit: $\alpha_{\rm bright}=-0.13 \pm 0.16$ and $\alpha_{\rm dim}=-0.40 \pm 0.36$, respectively). When we consider instead integrated fluxes (which have larger systematic errors due to fitting), we have $\alpha_{\rm dim}=0.2 \pm 0.7$ and $\alpha_{\rm bright} = -0.6\pm0.3$, but the latter is mostly driven by the first point at 4.55 GHz, and the one at 5.9 GHz (taking out these two points, the slope is consistent with being flat). The maps of the spectral index derived via {\tt mtmfs} deconvolution also show spectral indexes in the range $\alpha \sim (-0.4)$-0 in the source region. Overall, the spectra are compatible with being slightly negative, close to flat, in both bright and dim epochs. No clear spectral difference between the two flux states is seen. The relatively narrow band and the uncertainty on the slope prevents us from inferring physical quantities of the emitting electron population.

No epoch or combination of them shows signs of relevant circular polarisation in C-band. Taking individual observations, we can set 3$\sigma$ upper limits as low as $|V/I|\lesssim 7\%$ for the brightest epoch in C-band. Combining all epochs together ($\sigma\sim 1.4\, \mu$Jy/beam, average Stokes $I$ $\sim 80\, \mu$Jy/beam), or the seven brightest ones only ($\sigma\sim 2.6\, \mu$Jy/beam, average Stokes $I$ $\sim 170\, \mu$Jy/beam) provide in both cases a slightly stricter 3$\sigma$ upper limit $|V/I|\lesssim 5\%$.
The lack of circular polarisation, together with no significant sinusoidal-like modulation detected in the GLS periodogram at the orbital and synodic periods of each of the four planets, lead us to conclude that there is no evidence for magnetic star-planet interaction in our C-band data. 

\subsubsection{Correlation between radio and optical flux}\label{sec:corr_radio_optical}

\begin{figure}
    \centering
    \includegraphics[width=0.45\textwidth]{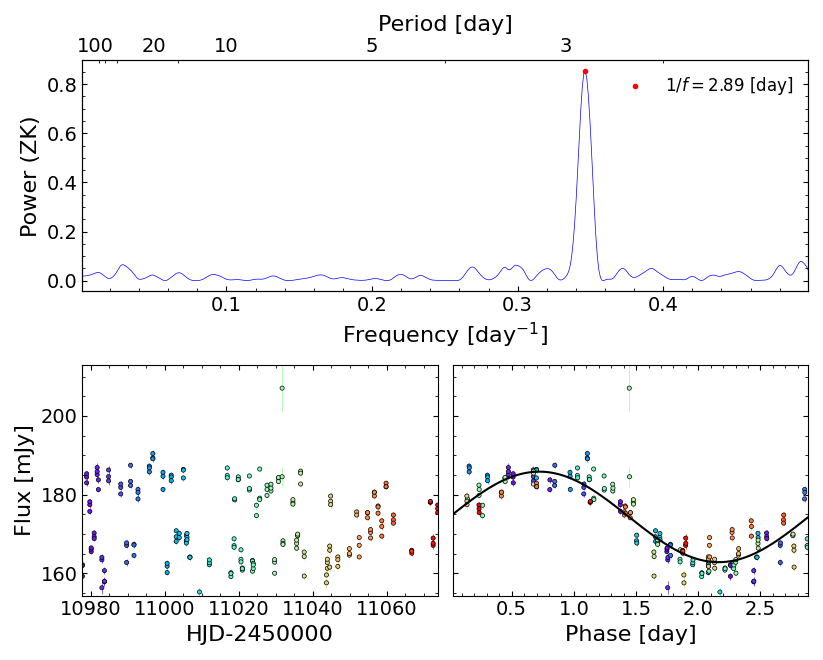}\\
    \caption{ASAS-SN optical data contemporary to the radio JVLA observations. \textit{Top:} GLS periodogram, showing the main peak at $P=2.89$ d, corresponding to the known stellar rotation period. \textit{Bottom left:} time series. \textit{Bottom right:} corresponding phase-folded light curve, with colours identifying the epoch (as in bottom left panel).}
    \label{fig:glsasas}
\end{figure}

\begin{figure}
    \centering
    \includegraphics[width=0.5\textwidth]{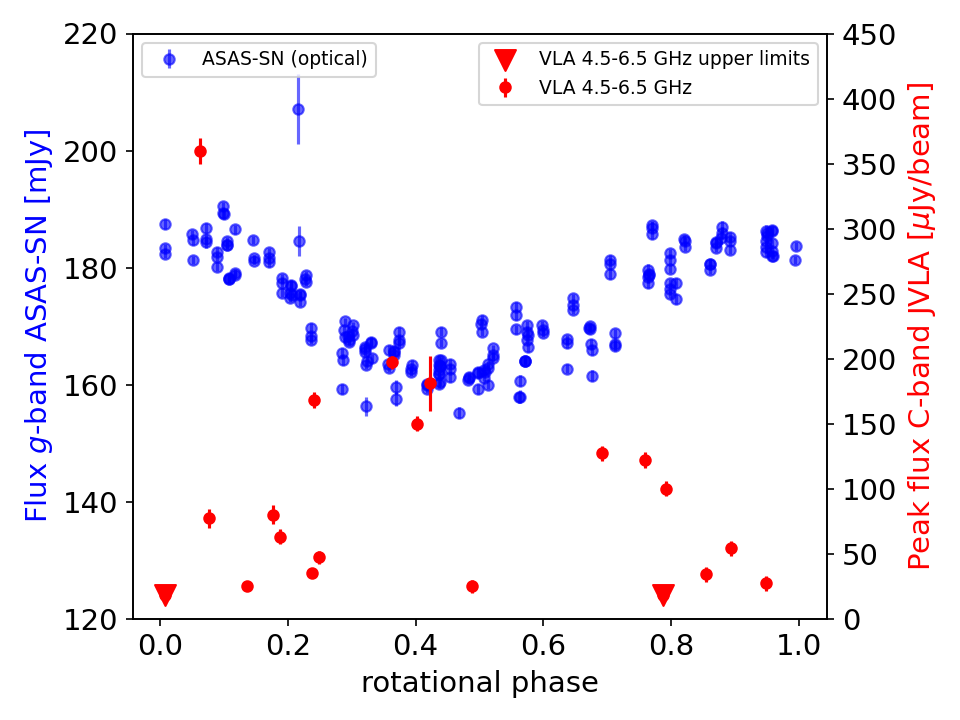}
    \caption{JVLA Band C peak flux and ASAS-SN optical flux collected within the same timespan, phase-folded to the stellar rotation period, as in Fig. \ref{fig:glsasas}. Red dots identify peak flux of radio detections, and red triangles correspond to 3$\sigma$ upper limits in peak flux (observations 2 and 3).}
    \label{fig:phase_folded}
\end{figure}

\begin{figure}
    \centering
    \includegraphics[width=0.5\textwidth]{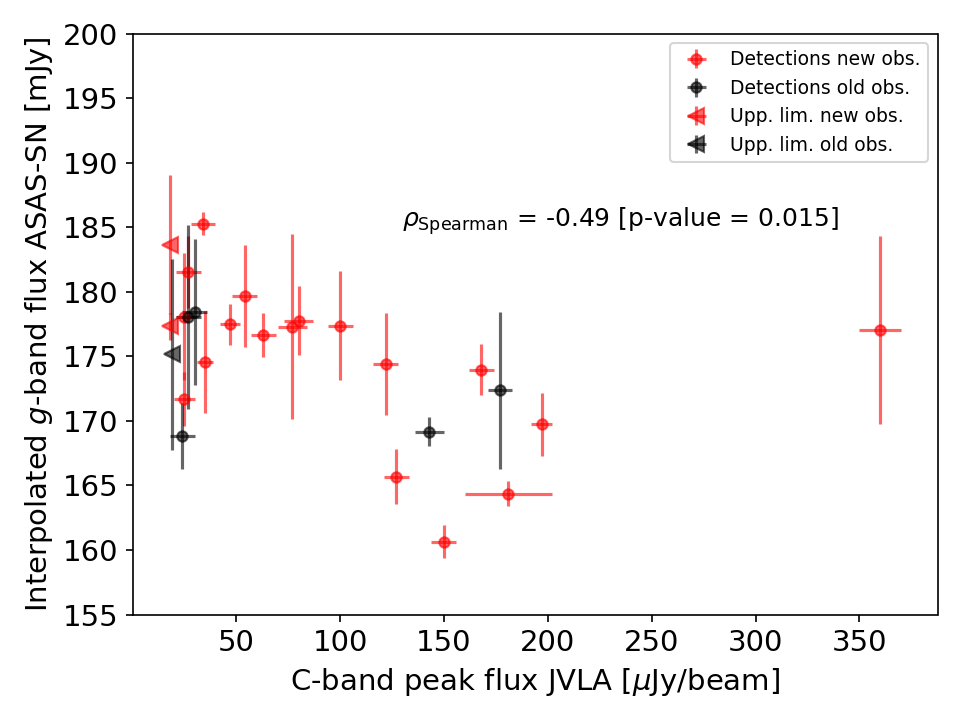}
    \caption{Correlation between optical and radio peak fluxes in Band C. As in Fig. \ref{fig:phase_folded}, red dots and triangles identify the detections and $3\sigma$ upper limits from the 2025-2026 campaign, respectively. Black dots and triangles correspond to detections and $3\sigma$ upper limits from \cite{damasso25}, respectively. The optical \textit{g}-band data are values interpolated at the epochs of JVLA observations. The reported Spearman's rank correlation coefficient corresponds to joint old and new data, excluding C-band upper limits.}
    \label{fig:correlation}
\end{figure}

Following the analysis in D25, we used the ASAS-SN light curve contemporary to the time span covered by JVLA data (from 2460977.678 to 2461073.786 HJD) to monitor the time variability of the optical flux, which is known to be modulated by the star’s rotation period, and investigate whether it is correlated with emissions in band C. This behaviour was tentatively identified by D25 with fewer radio data points available, suggestive that the variability observed at radio frequencies was linked to stellar magnetic activity.
Figure \ref{fig:glsasas} shows the GLS periodogram of the ASAS-SN optical light curve. The main peak occurs at $P=2.89$ d, compatible with the known $2.91\pm0.05$ d stellar rotation period \citep[e.g. ][]{Suarez2022NatAs...6..232S}. We phase-folded ASAS-SN and JVLA measurements\footnote{The epochs of radio and X-ray/NUV observations are expressed in MJD, and were transformed to the HJD reference to be consistent with the optical measurements.} using this period (Fig. \ref{fig:phase_folded}). The comparative plot tentatively confirms the finding of D25 that the highest radio peak fluxes tend to occur during phases of optical minimum in the ASAS-SN data, which arguably correspond to epochs of maximum coverage of the photosphere by darker and magnetically active regions.  Figure \ref{fig:correlation} includes the C-band fluxes reported by D25 and this work (26 points), with the corresponding Gaussian process (GP) optical flux predictions. To perform this correlation analysis we have interpolated the ASAS-SN light curve to the JVLA epochs by modelling the optical flux through a GP regression, by leveraging the GP's high predictive power for a time series that accurately samples the stellar rotation period through intensive monitoring (see App. \ref{sec:gpasas} for details). The plot supports evidence of a moderate anti-correlation between optical and radio peak fluxes shown in Fig. \ref{fig:correlation}, with a Spearman's rank correlation coefficient (excluding the two radio upper limits) $\rho_{\rm Spearman}=-0.49$, with a corresponding p-value of 1.5\%, which is indicative of an anti-correlation, but still leaves the door open to mere statistical fluctuations.

The epoch of the brightest radio emission (observation  7, with flux 360$\pm$10 $\mu$Jy/beam) deviates from the trend. However, we don't have a clear reason to label it as an outlier. It could be if, for instance, the flux states were rotation-modulated while that epoch was due to a transient intense burst, unrelated to rotation. However, we don't have simultaneous measurements (e.g., showing an optical flare) to support this, since the optical flux at the time of each radio observation is reconstructed using GP interpolation of ASAS-SN data. More importantly, no short-term radio variability within the duration of the visit is observed, thus discarding a short (minutes), very intense radio burst as responsible for the high flux. Similarly, one dim epoch (around phase 0.5) deviates from the trend. The presence of low radio states within the rotational phase range where high emissions have been seen in AU Mic \citep[while no high state is seen outside such range][]{aumic24}. The putative anti-correlation spoiled by these two data points make the interpretation difficult, and call for a much larger data sample.

In App. \ref{sec:spot_modeling} we describe a complementary correlation analysis based on the modelling of the optical light curve to reconstruct the stellar surface inhomogeneities at the time of radio observations. When excluding observation 7 from the analysis, we found a moderate positive correlation between radio emission and projected stellar spot filling factor, in agreement with the optical-radio possible anti-correlation. This might support the scenario that the observed radio emission is linked to the presence of active regions and, consequently, to stellar magnetic activity. 

\subsection{X-ray and NUV emission}
The results of the multi-epoch analysis are summarized in Table \ref{tableswift}, where we report the Near-Ultraviolet (NUV) luminosity, the plasma temperature ($kT$), the emission measure ($EM$), and the X-ray luminosity ($L_X$) for each observation. The time series of the X/NUV luminosities is shown in the first panel of Fig. \ref{fig:jvlatimeseries}.

\begin{figure}
    \centering
  \includegraphics[width=0.45\textwidth]{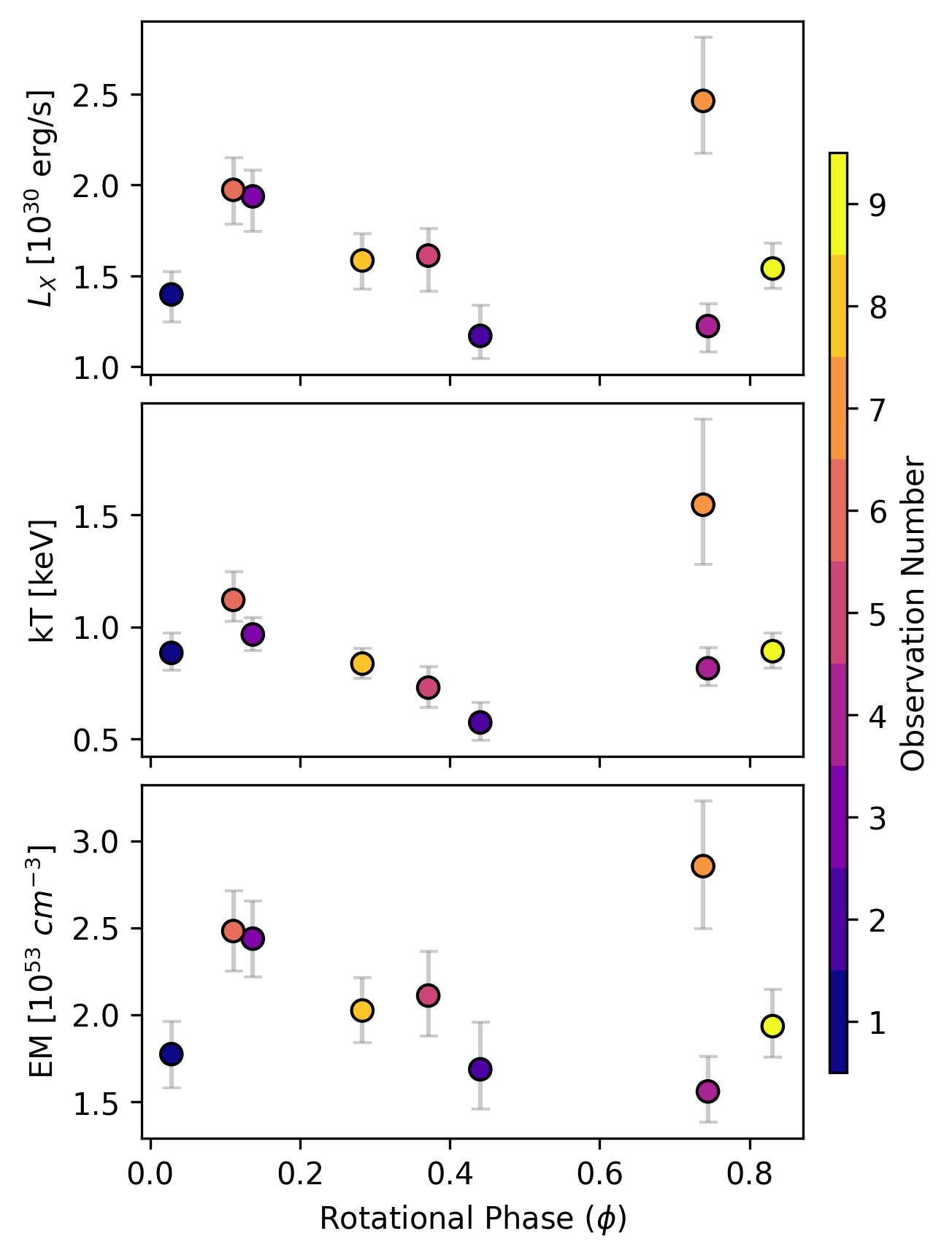}
    \caption{X-ray parameters of V1298\,Tau phase-folded with the stellar rotational period ($P_{\rm rot}$=2.89~d). From top to bottom: X-ray luminosity ($L_{\rm X}$), coronal temperature ($kT$) and emission measure ($EM$). The data points are color-coded according to the observation sequence (1--9), as indicated by the color bar on the right. Vertical bars represent the 1$\sigma$ uncertainties.}
    \label{xraynuvanalysis}
\end{figure}

The NUV luminosity, $L_{NUV}$, shows only modest variability, ranging from a minimum of $(1.63 \pm 0.05) \times 10^{31}$ erg s$^{-1}$ (observation 5) to a maximum of $(1.81 \pm 0.05) \times 10^{31}$ erg s$^{-1}$ (observation 3). Regarding the X-ray spectral properties, the coronal temperature $kT$ fluctuates between $0.57_{-0.08}^{+0.09}$ keV and $1.54_{-0.27}^{+0.38}$ keV, while the emission measure $EM$ varies within the range $[1.56 - 2.85] \times 10^{53}$ cm$^{-3}$. Consequently, the X-ray luminosity $L_X$ exhibits significant changes, spanning from $1.17_{-0.13}^{+0.17} \times 10^{30}$ erg s$^{-1}$ to $2.46_{-0.29}^{+0.35} \times 10^{30}$ erg s$^{-1}$. In Fig. \ref{xraynuvanalysis}, we show the temporal evolution of these parameters folded as a function of the stellar rotational period $P=2.89$~d (see above). The current dataset is insufficient to look for periodicity, and shows no clear evidence for rotational modulation or modulation with planetary periods.

To investigate the X-ray variability of V1298\,Tau, we also performed a systematic analysis of the Swift-XRT light curves. To isolate the quiescent emission level and identify transient events, we employed an iterative sigma-clipping algorithm to calculate the robust median and the standard deviation ($\sigma$) of the count rate, effectively filtering out outlying data points associated with flaring activity. A flare candidate was statistically defined as a sequence of at least two consecutive bins exceeding the $3\sigma$ threshold above the clipped median. This stringent criterion was adopted to minimize the impact of Poissonian noise and potential instrumental artifacts, such as cosmic ray strikes, ensuring that only physically motivated temporal evolutions - consistent with the rapid rise and decay phases of stellar magnetic reconnection events - were classified as flares. The variability analysis revealed the presence of a single, prominent flare occurring at the onset of the first observation. Specifically, the light curve captures only the decay phase of the event, with the initial rise and peak occurring prior to the start of the XRT exposure (Fig. \ref{fig:flare}). No other significant flaring events were detected in the subsequent observations, which remained consistent with a stable quiescent state within the sensitivity limits of our analysis.

\begin{table*}[h]
\centering
\tiny
\caption{Results of the Swift observations in X-ray and NUV for V1298\,Tau.}
\begin{tabular}{ccccccccc}
\hline\hline
Obs. & obsid & Obs. Start & MJD-2400000 & Exp. & $L_{NUV}$  & $kT$ & $EM$ & $L_{X}$ \\
 & & [yyyy-mm-dd hh:mm:ss] & [day] & [s] & [$10^{31}$ erg/s] & [keV] &  [$10^{53}$ cm$^{-3}$] & [$10^{30}$ erg/s] \\
\hline
1 & 03000068001 & 2025-09-10 16:09:00 & 60928.780 & 2680 & $1.71 \pm 0.05$ & $0.88_{-0.08}^{+0.09}$ & $1.77 \pm 0.19$ & $1.40_{-0.15}^{+0.13}$ \\
2 & 03000068002 & 2025-09-14 19:38:00 & 60932.860 & 2730 & $1.66 \pm 0.04$ & $0.57_{-0.08}^{+0.09}$ & $1.69_{-0.23}^{+0.27}$ & $1.17_{-0.13}^{+0.17}$ \\
3 & 03000068003 & 2025-09-16 19:50:00 & 60934.872 & 2885 & $1.81 \pm 0.05$ & $0.97_{-0.07}^{+0.08}$ & $2.44_{-0.22}^{+0.21}$ & $1.94_{-0.19}^{+0.15}$ \\
4 & 03000068004 & 2025-09-18 10:59:00 & 60936.627 & 2440 & $1.64 \pm 0.07$ & $0.82_{-0.08}^{+0.09}$ & $1.56_{-0.18}^{+0.20}$ & $1.22_{-0.14}^{+0.12}$ \\
5 & 03000068005 & 2025-09-20 09:25:00 & 60938.437 & 2230 & $1.63 \pm 0.05$ & $0.73_{-0.09}^{+0.10}$ & $2.11_{-0.23}^{+0.26}$ & $1.61_{-0.20}^{+0.15}$ \\
6 & 03000068006 & 2025-09-22 08:09:00 & 60940.574 & 2420 & $1.69 \pm 0.05$ & $1.12_{-0.10}^{+0.13}$ & $2.48 \pm 0.23$ & $1.97_{-0.19}^{+0.18}$ \\
7 & 03000068007 & 2025-09-24 08:12:00 & 60942.387 & 1360 & $1.79 \pm 0.07$ & $1.54_{-0.27}^{+0.38}$ & $2.85_{-0.36}^{+0.38}$ & $2.46_{-0.29}^{+0.35}$ \\
8 & 03000068009 & 2025-09-28 18:01:00 & 60946.850 & 2770 & $1.73 \pm 0.05$ & $0.84_{-0.06}^{+0.07}$ & $2.02 \pm 0.19$ & $1.58_{-0.16}^{+0.15}$ \\
9 & 03000068010 & 2025-09-30 03:52:00 & 60948.433 & 2620 & $1.67 \pm 0.05$ & $0.89_{-0.07}^{+0.08}$ & $1.93_{-0.18}^{+0.21}$ & $1.54_{-0.11}^{+0.14}$ \\
\hline\hline
\end{tabular}
\tablefoot{\tiny
{Log of the Swift observations of V1298\,Tau during the September 2025 campaign. The table lists the observation ID, start time, effective exposure time, and NUV luminosity (UVM2 filter). X-ray spectral parameters ($kT_1$, $EM_1$, and $L_X$) were derived from a single-temperature (1T) APEC model in the 0.3–10 keV energy range, assuming a distance of 108.02 pc. Interstellar absorption and metallicity were fixed to the values described in Section \ref{sec:X}. Uncertainties are quoted at the 68\% confidence level.}
}
\label{tableswift}
\end{table*}

\begin{figure}
    \centering
    \includegraphics[width=0.45\textwidth]{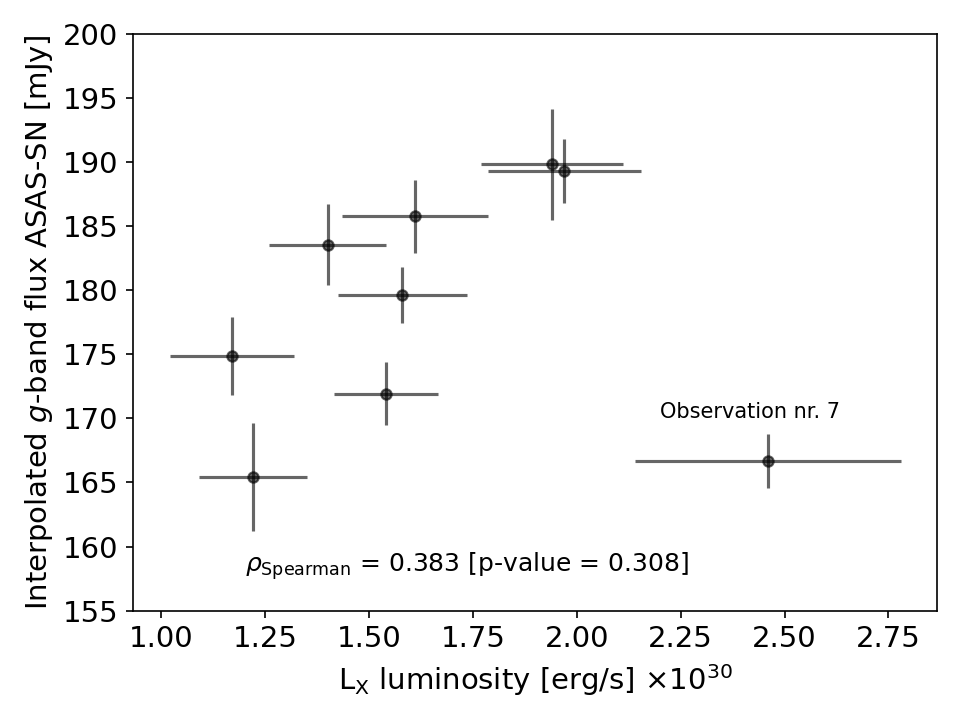}\\
    \includegraphics[width=0.45\textwidth]{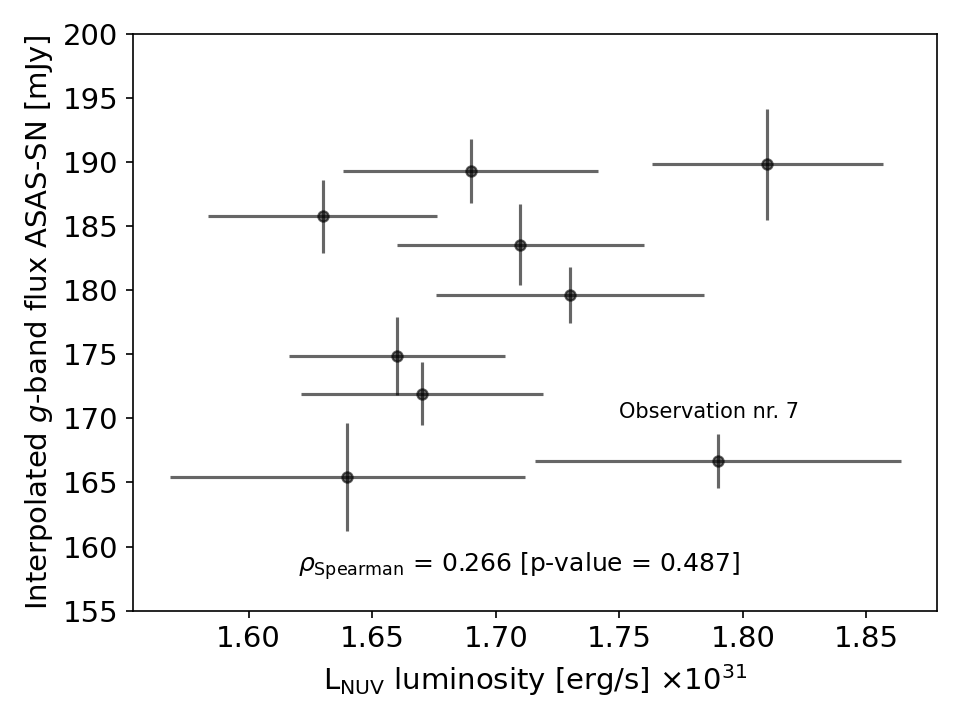}
    \caption{Correlations between the X-ray/NUV luminosities and the ASAS-SN optical fluxes predicted at the epochs of Swift observations through a GP regression. \textit{First panel.} X-ray luminosities. \textit{Second panel.} NUV luminosities.}
    \label{fig:lxnuvasas}
\end{figure}

\begin{figure}
    \centering
  \includegraphics[width=0.45\textwidth]{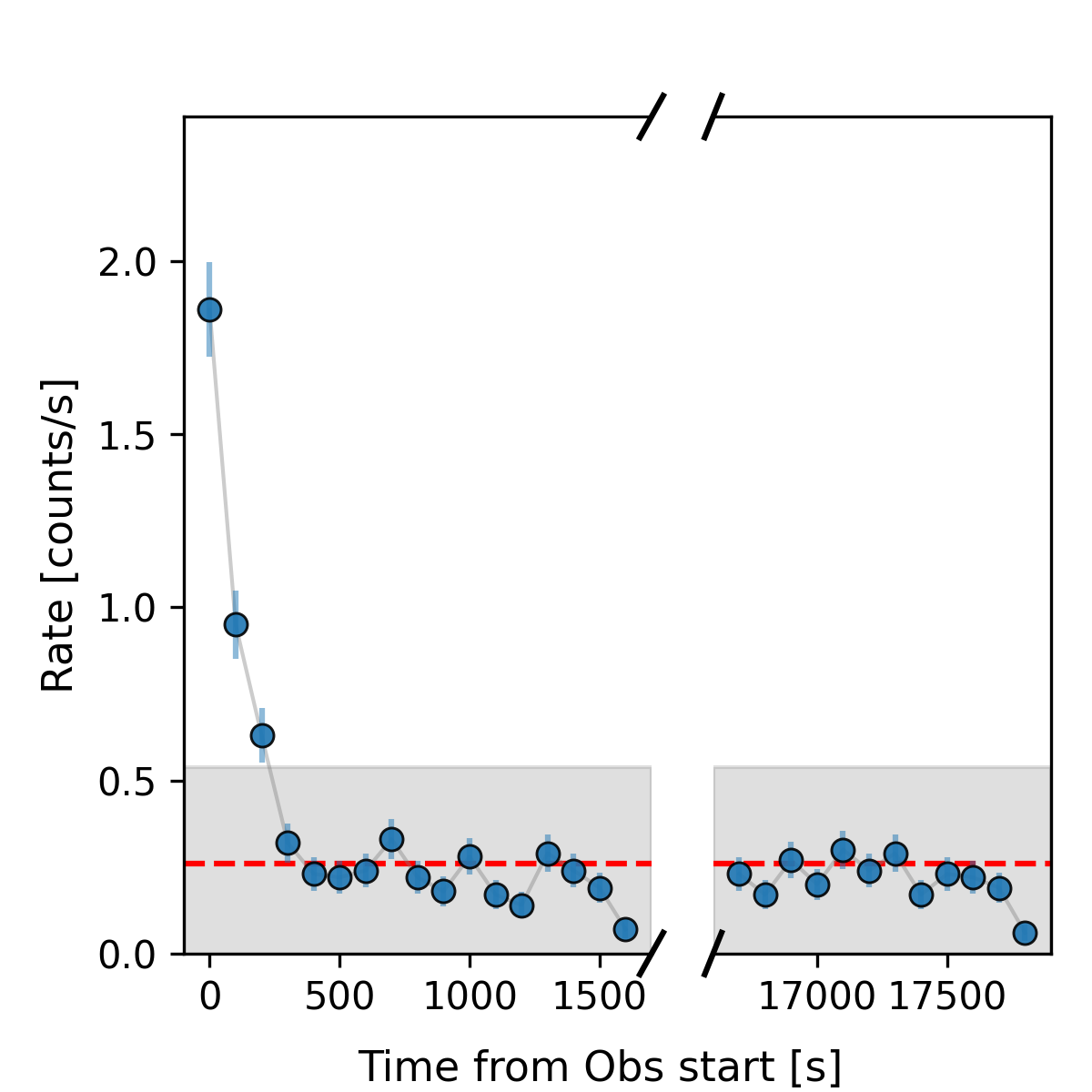}
    \caption{Swift-XRT X-ray light curve (0.3--10 keV) of V1298\,Tau during the first observation (Obs. 1). The data shows the decay phase of a prominent stellar flare. The horizontal dashed line represents the quiescent median level determined through iterative $\sigma$-clipping, while the dotted line indicates the $3\sigma$ threshold used for flare identification. The event began prior to the start of the XRT exposure, leaving only the cooling phase visible. Error bars represent $1\sigma$ Poissonian uncertainties}
    \label{fig:flare}
\end{figure}

\subsection{X-ray, NUV, and radio luminosity correlations} \label{sec:multiwavecorrelation}
To investigate the physical connection between the coronal and chromospheric/transition region activity, we evaluated the relationship between the X-ray ($L_{\rm X}$) and near-ultraviolet ($L_{\rm NUV}$) luminosities. An analysis based on Spearman's correlation yields a coefficient of $\rho_{\rm Spearman}$ = +0.53 with a p-value of 0.14, indicating a statistically moderate positive correlation. This would suggest that the emission in both energy regimes could be physically coupled, likely originating from the same magnetic structures and activity phenomena. 

To investigate the interplay between the high-energy atmospheric layers and the underlying photosphere, we analysed the correlation between the X-ray and NUV luminosities and the simultaneous optical $g$-band fluxes, the latter interpolated at the Swift epochs via GP regression (Fig.\ref{fig:lxnuvasas}). The Spearman's rank correlation coefficients indicate weak correlations ($\rho_{\text{Spearman}} = +0.38$ and $+0.27$ for $L_{\text{X}}$ and $L_{\text{NUV}}$, respectively, both with high p-values). Clearly, the result of this analysis is strongly influenced by the low number of available data, nonetheless we note that it points out a possible positive correlation, especially between optical flux and X-ray luminosity (top panel)\footnote{The contrast between the radio and X-ray correlations with optical flux could be reflect their differing emission geometries and optical depths. The C-band radio flux peaks during optical minima because the gyrosynchrotron and/or cyclotron emission comes from intense coronal magnetic fields localised in photospheric dark spots. Conversely, X-ray emission is optically thin and traces a much larger coronal volume. In active stars, tall coronal loops originating from high-latitude or far-side active regions can extend past the limb, therefore this decoupled geometry allows the X-ray flux to remain high independently of the concurrent, line-of-sight dark spot coverage modulating the optical flux.}, although observation 7 outlies such a putative trend. If we exclude observation 7 from the dataset, we get $\rho_{\text{Spearman}}$=+0.81 (p-value 0.015) and $\rho_{\text{Spearman}}$=+0.48 (p-value 0.23), for the X-ray and NUV luminosities, respectively, although it is not straightforward to justify this data point as an outlier. During this visit of the Swift satellite, the highest $L_{\text{X}}$ and elevated $L_{\text{NUV}}$ values recorded during the campaign coincide with a predicted low optical flux. This suggests that the intense high-energy emission was registered during a phase of optical minimum, corresponding to a configuration of maximum photospheric coverage by dark, magnetically active starspots. Crucially, the X-ray count rate during this specific epoch does not exhibit a classic, sharp impulsive flare profile. This suggests that observation 7 either captured the sustained plateau phase of a long-duration flare, where high-energy emission persists after the optical and ultraviolet continuum has subsided \citep[e.g.][]{gudel2002, pillitteri2022, kowalski2024}, or a dense coronal active region during a phase characterised by low photospheric flux due to an increased starspot coverage.

In general, the correlation seen in Fig.~\ref{fig:lxnuvasas} is low suggesting that the NUV emission in the UVM2 filter passband of Swift, covering a range centered around 225~nm with a FWHM of about 50~nm, is only weakly associated with dark starspots in the active regions of V~1298~Tau. The main source of the continuum in that band is the bound-free emission of ionized hydrogen that recombines to the quantum level $n=2$, thus emitting radiation below the Balmer jump at 364.5~nm. The sources of hydrogen ionizing radiation are not strictly associated with dark spots, therefore, the ensuing bound-free radiation from hydrogen recombination is not expected to be strictly connected with the location of starspots. On the other hand, the microwave flux at 5 GHz has a relevant component that comes from electron gyrosynchrotron emission in the intense magnetic fields of starspots as suggested by the strong association of the solar 2.8 GHz emission (the so-called 10.7-cm radiation) with sunspots and the low corona of active regions \citep[see][for details]{TappingDetracey90}. 

\begin{figure}
    \centering
  \includegraphics[width=0.45\textwidth]{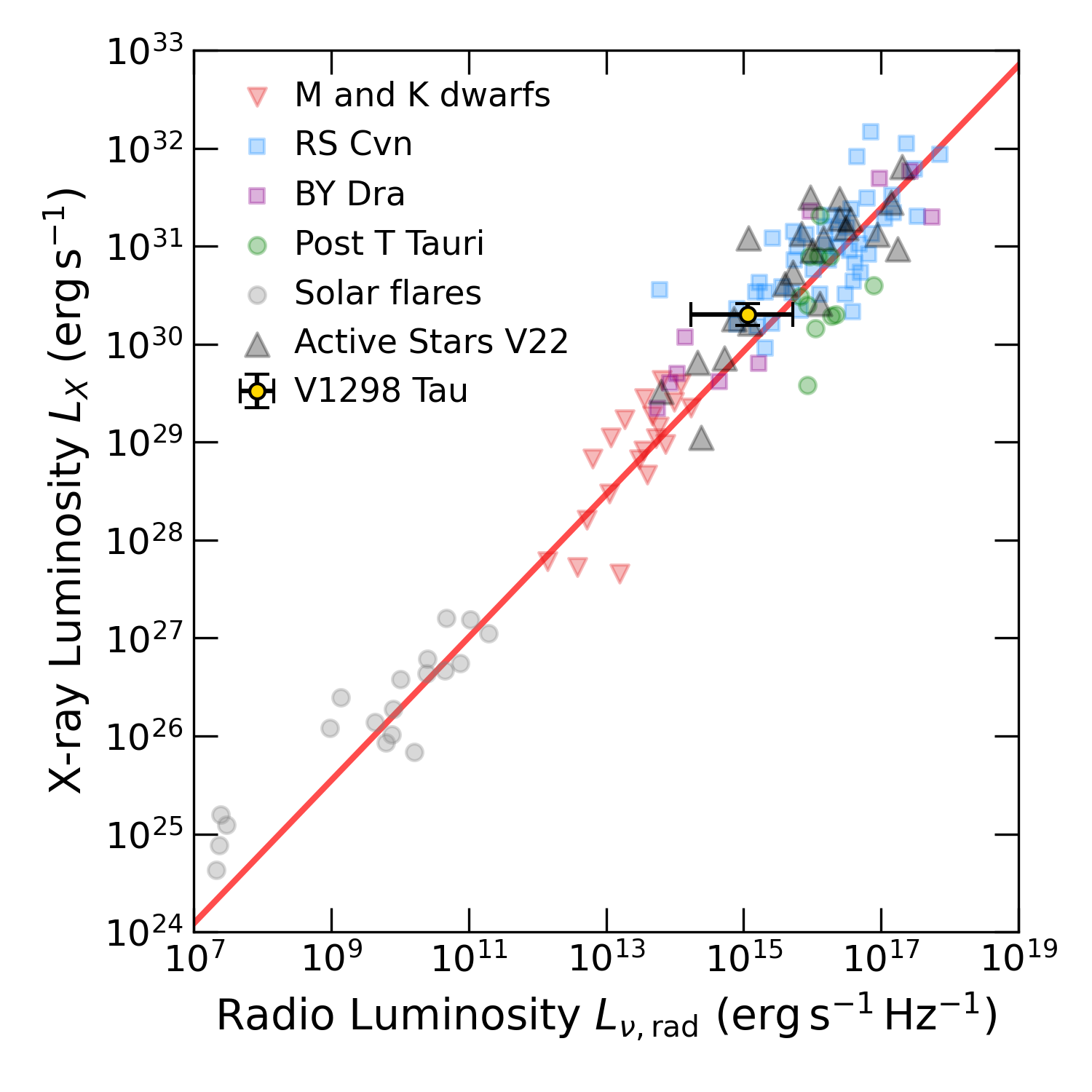}
    \caption{X-ray luminosity ($L_{\rm X}$ in the 0.1–2.4 keV band) as a function of the radio luminosity ($L_{\nu, \rm rad}$) for V1298\,Tau (yellow circle) compared to various classes of magnetically active sources. The solid red line represents the Güdel-Benz relation $L_{X} = 9.48 \times 10^{18} L_{\nu, rad}^{0.73}$ \citep{Williams2014}. Comparison samples, including solar flares (grey circles), M and K dwarfs (pink triangles), RS CVn systems (blue squares), BY Dra variables (purple squares), and Post T Tauri stars (green circles), are taken from \cite{Vedantham2022}. For V1298\,Tau, the data point indicates the median luminosity, while the error bars denote the observed range between the minimum and maximum values across the observing campaign.}
    \label{fig:gudel}
\end{figure}

Furthermore, we calculated the X-ray luminosity ($L_{\rm X}$) in the 0.1–2.4 keV band by extrapolating the best-fit spectra from the XRT data, aiming to assess whether the source aligns with the Güdel-Benz relation \citep{Guedel1993}. To this end, we computed the median values of both the 0.1–2.4 keV X-ray luminosity and the radio luminosity when the source is detected, adopting the minimum and maximum observed values as the lower and upper uncertainties, respectively. Figure \ref{fig:gudel} displays the Güdel-Benz relation alongside various classes of objects taken from \cite{Vedantham2022}. As is evident, the source lies perfectly on the relation within the estimated errors. Specifically, the observed X-ray and radio luminosity places V1298\,Tau in the high-luminosity regime, consistent with the cluster of Post T Tauri stars and active RS CVn systems. This position reinforces the status of V1298\,Tau as a young, magnetically active pre-main sequence star, where the physical mechanisms coupling coronal heating and non-thermal particle acceleration remain consistent with those of other active stellar populations. 

\section{Discussion and conclusions} \label{sec:discussion}

We presented results from a multi-wavelength follow-up campaign of the infant star V1298\,Tau which is host to a multi-planetary system. The monitoring covered the X-ray and NUV bands for twenty days in September 2025 using the Swift satellite, and the radio band (C-band, 4.5–6.5 GHz) between late October 2025 and early February 2026 using JVLA. The radio and high-energy campaign was not simultaneous, nonetheless it allows for a characterisation of stellar activity variability at different wavelengths over a time span of nearly five months, supported by the optical light curve from the ASAS-SN survey.  

Concerning radio frequencies, we detected emission above the 3$\sigma=6$\,$\mu$Jy/beam in 90\% of the visits, with peak fluxes in the range $\sim25-360$ $\mu$Jy/beam. This result confirms that V1298\,Tau is a radio emitter, as found by D25 with a lower number of epochs, and it significantly reinforces the finding of D25 that this emission is correlated with the variability observed in the optical band, based on a comparison with the light curve obtained by ASAS-SN during the same observation period. Spectral analysis reveals a flat or slightly negative frequency slope. Similarly to D25, the detected emissions do not show signs of circular polarisation, with $|V/I|\lesssim 5\%$. The lack of relevant circular polarisation, the persistence of the emission, and the hours/days variability timescales would suggest that, similarly as concluded in D25, we detected in Band C gyrosynchrotron and/or cyclotron emission from the corona related to stellar magnetic activity.

The variability, and, more specifically, close-spaced observations collected in January 2026 are suggestive of a jump between different states, maybe in form of long-lived flare event spanning several days (Fig. \ref{fig:jvlatimeseries}). A combination of flares with shorter decay time is also possible, because multiple spikes have been observed, but short-term variability is only seen in the last observation (a one-order-of-magnitude flux rise within minutes). The similar but younger ($\sim$2 Myr) system V\,830\,Tau shows qualitatively resemblant behaviour at $4-8$ GHz, although at higher fluxes ($\sim 0.3-1.2$ mJy): time variability of radio flux over hours or days, slightly negative spectral slope, small to undetectable circular polarisation \citep{osten25}. These authors put forward the hypothesis that it could be triggered by the presence of a controversial close-in planet candidate, in a down-scale version of the interaction seen in tight binary star (e.g. RS CVn, BY Dra systems), where also radio emission tend to show orbital modulation or enhancement around periastron \citep{richards03,massi06,salter10}. In the future, it would be interesting to test this suggestive scenario for the confirmed compact multi-planet system V1298\,Tau by collecting more data.
Simulations to reconstruct stellar photospheric inhomogeneities, in the form of spots and faculae, at the time of the radio follow-up show potential moderate-to-strong correlation between the radio emission and the projected filling factor of the active regions.  

The high-energy monitoring with Swift revealed minor and relevant variability in the NUV and X-ray luminosity, respectively, and the final stage of an intense flare was recorded in X-rays. Our data confirm that V1298\,Tau follows the Güdel-Benz relation, with X-ray and radio luminosities consistent with those of magnetically active pre-main sequence star. 
 
\begin{acknowledgements}
We thank the anonymous referee for their constructive report. This work is based on observations carried out with the Karl G. Jansky Very Large Array (project 25B-133), and with the Swift satellite. DV is supported by the European Research Council under the European Union’s Horizon 2020 research and innovation programme (ERC Starting Grant "IMAGINE" No. 948582). 
We acknowledge the ``Mar\'ia de Maeztu'' award from the Spanish Science and Innovation Ministry to the Institut de Ci\`encies de l'Espai (CEX2020-001058-M) and the MaX-CSIC Excellence Award MaX4-SOMMA-ICE, by the Generalitat de Catalunya/CERCA programme.
A.S.-M.\ acknowledges support from the RyC2021-032892-I grant funded by MCIN/AEI/10.13039/501100011033 and by the European Union `Next GenerationEU’/PRTR. JMG acknowledges financial support through the grant PID2023-146675NB-I00.
Views and opinions expressed are however those of the author(s) only and do not necessarily reflect those of the European Union or the European Research Council. Neither the European Union nor the granting authority can be held responsible for them. RS acknowledges the support of the ARIEL ASI/INAF agreement 2021-5-HH.0 plus addenda no. 2021-5-HH.1-2022 and 2021-5-HH.2-2024 and the support from the European Union – Next Generation EU through the grant n. 2022J7ZFRA – Exo-planetary Cloudy Atmospheres and Stellar High energy (Exo-CASH) funded by MUR – PRIN 2022. This work has been partially supported by the ASI-INAF program I/004/11/6. SK acknowledges financial support by the Generalitat Valenciana (Spain) through the PROMETEO Programme for Excellence Research Groups under grant CIPROM/2022/64.
\end{acknowledgements}

%
\bibliographystyle{aa} 
\bibliography{v1298tauradio} 
%

\appendix

\section{A closer look to radio maps}\label{app:maps}

\begin{figure*}
\centering
\setlength{\tabcolsep}{6pt} 

\newcommand{\panelH}[3][7.6cm]{
  \begin{tikzpicture}[baseline=(img.north)]
    \node[inner sep=0, outer sep=0] (img)
      {\includegraphics[height=#1, keepaspectratio]{#2}};
    \node[anchor=north, font=\small, text=black,
          fill=white, rounded corners=1pt, inner sep=1pt,
          fill opacity=0.6, text opacity=1,
          yshift=-4pt] at (img.north) {#3};
  \end{tikzpicture}%
}

\scalebox{0.95}{
\begin{tabular}{@{}p{0.485\textwidth} p{0.485\textwidth}@{}}
\centering
\panelH[7.6cm]{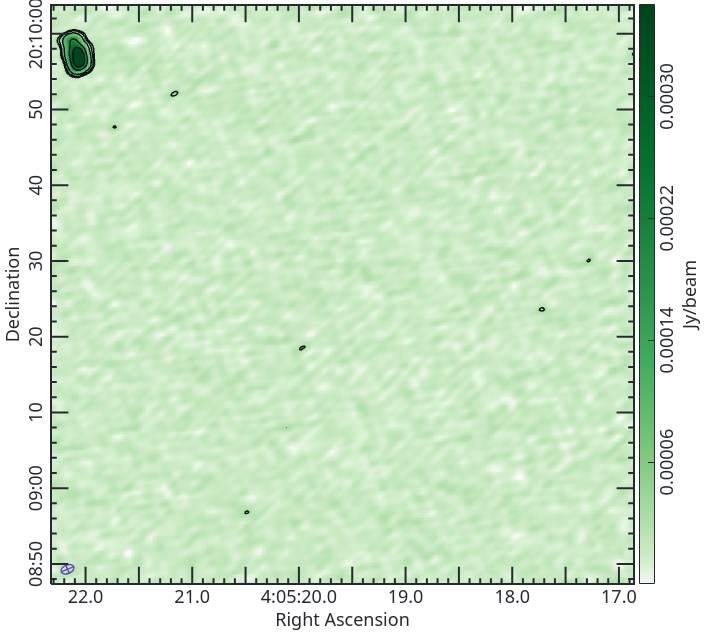}{Observation 2} &
\panelH[7.6cm]{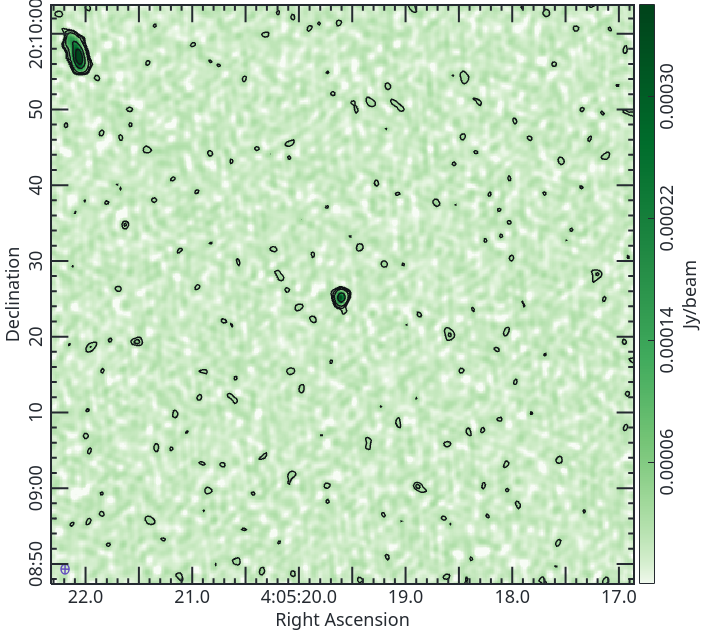}{Observation 7} \\
\centering
\panelH[7.6cm]{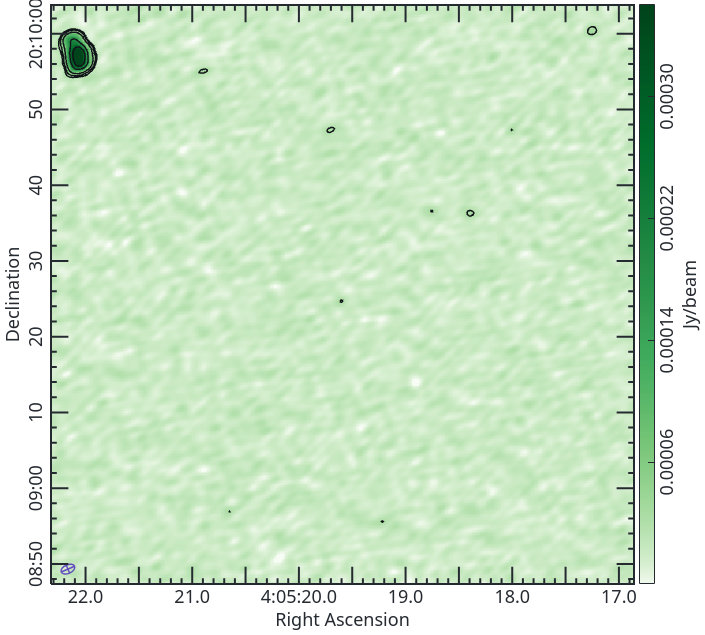}{Observation 13} &
\panelH[7.6cm]{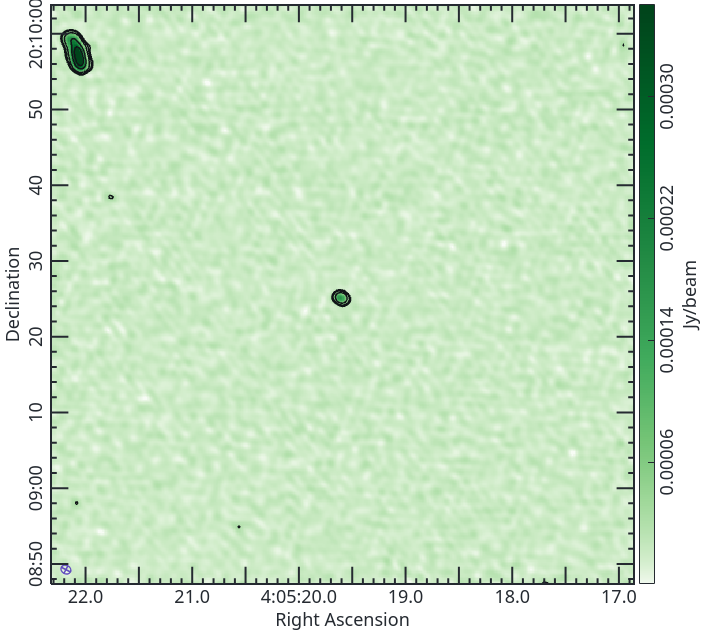}{Observation 14} \\
\centering
\panelH[7.6cm]{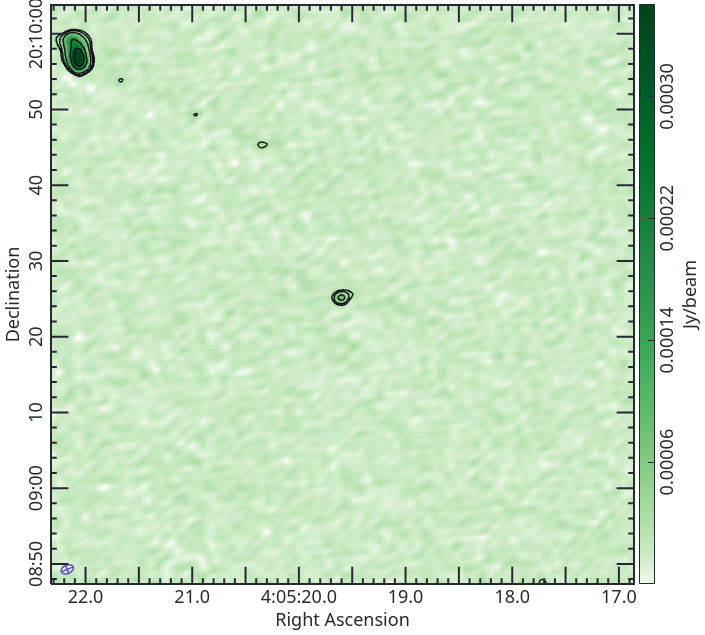}{Observation 16} &
\panelH[7.6cm]{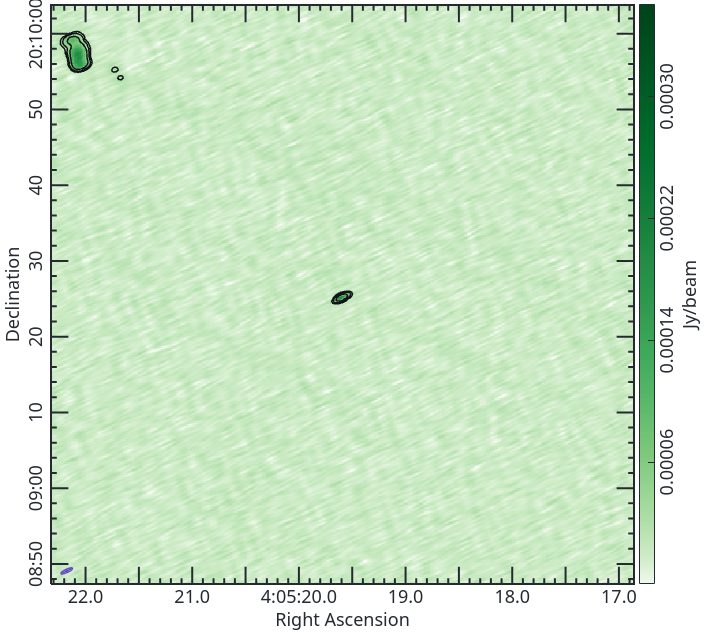}{Observation 20}
\end{tabular}
}
\vspace{-0.25\baselineskip}
\caption{Sample of clean Stokes I images, centered on V1298 Tau's position, for six representative epochs. From top left to bottom right, the observations ID from Table \ref{tab:radioresults} are: 2 (non-detection), 7 (brightest day and higher rms than usual), 13 (barely detected), 14 (strong detection a few hours later observation 13), 16 (intermediate flux), 20 (second brightest day, with short-term variations, see Fig. \ref{fig:radio_variations}, and elongated beam due to the BnA configuration). The beam is indicated in purple in the bottom left corner. The common color scale indicate the flux density, and black contours indicate (3, 5, 10, 30 and 50) $\times$ 6 $\mu$Jy/beam flux levels.}
\label{fig:images}
\end{figure*}

Figure \ref{fig:images} shows the clean images for a representative sample of JVLA observations, including the brightest epoch (top right), non-detection (top left), and the last day where variations are seen (bottom right).
There is an extended source, located at about 47 arcsec North-East to our source (top right corner in the maps), with a peak flux $\sim$ 0.3-0.4 mJy/beam, comparable to the brightest V1298\,Tau's measured flux. It is catalogued as the radio-source NVSS J040522+200958 (RA: 04h05m22.07s, Dec.: +20:09:58.3), with no known counterparts according to Simbad database. Other three background sources with similar fluxes ($\sim$0.2-0.5 mJy/beam) are located within the primary beam, but much further away (3-5 arcmin).

\section{Nearby and common proper motion sources to V1298\,Tau} \label{app:nearbysources}
\subsection{Radio observations}
Given that this field has never been observed by dedicated (sub)-GHz observations (by e.g. VLA, GMRT, LOFAR), we also looked for the emission of two interesting sources. The first one is HD\,284154 (RA 04h05m14.352s, Dec +20d08'21.542"), a double-lined spectroscopic binary \citep[distance 107.65$\pm$0.27 pc and orbital period 17.75 d][]{gaia2023A&A...674A...1G,2023A&A...674A..34G} and physical companion to V1298\,Tau \citep[projected on-sky separation of 97.7 arcsec, corresponding to $\sim$10\,600 au at their distance][]{Suarez2022NatAs...6..232S}. The binary is coeval to V1298\,Tau, and the two components are similar early G-type stars. Our JVLA observations, also including those from D25, do not show detections at the pair location, at any epoch. Combining the 20 epochs of this work, we can put a $\lesssim 5$\, $\mu$Jy/beam 3$\sigma$ upper limit for the Stokes I emission in C-band.

The second object in the field of view is the candidate K-type young stellar object ATO J061.3032+20.1301 (RA 04h05m12.775s, Dec +20d07' 48.843"). It has been classified as young star by \cite{2021MNRAS.503..938Z}, and the parallax and proper motions from Gaia DR3 \citep[Gaia ID 51884824140206720][]{gaia2023A&A...674A...1G} indicate that it is a common proper motion companion to V1298\,Tau and HD\,284154. The large astrometric renormalised unit weight error (RUWE) of 1.66, and the large amplitude of the radial velocity range from Gaia DR3 $\sim$75 km/s) are consistent with a multiple stellar system. 
When combining all images (rms $\sim 1.5$\,$\mu$Jy/beam), we detect the source with Stokes I peak flux of $7\pm2$\,$\mu$Jy/beam (S/N$\sim$ 4-5), and no Stokes V (meaning, a 3$\sigma$ upper limit $|V/I|\lesssim 70\%$). The average flux level is comparable to the rms of individual epochs, and indeed we possibly detect this source only when it shows higher fluxes: in observation 2, with a possible extended, fully negatively circularly polarised emission, having peak flux of $\sim 21\pm7$\,$\mu$Jy/beam (S/N $\sim$ 3.5), integrated flux $50\pm 19$\,$\mu$Jy, and, even more tentatively, in observation 14, with peak flux $\sim 15\pm $\,$\mu$Jy/beam (S/N $\sim$ 2.5). Based on this minimal information, the young stellar object might show time variability from one epoch to another ascribable to polarised transients, but, given the barely detectable levels, and the lack of further dedicated observations in e.g. VLA, LOFAR, and GMRT archives, we cannot characterise it further. 

\subsection{X-ray spectral analysis}
Both targets HD\,284154 and ATO\,J061.3032+20.1301 fall in the field of view of Swift, and are detected in X-rays in our data, with detection by XMM-Newton EPIC reported in \cite{Maggio2023}. We co-added all individual \textit{Swift}/XRT observations into a single summed dataset. We modelled their combined spectra using a single-temperature (\texttt{1T}) \texttt{tbabs $\times$ apec} thermal model, maintaining $N_{\rm H} = 1.6 \times 10^{20}$\,cm$^{-2}$ and metallicity $Z = 0.2\,Z_{\odot}$ fixed to the same values. 

Because the two targets are separated by only $\approx 35''$, the point-spread function (PSF) of \textit{Swift}/XRT leads to mutual spatial cross-contamination within standard extraction apertures. Therefore, the quantities derived below reflect the total flux extracted from each region and must be considered conservative upper limits on the intrinsic emission of each target.

For HD\,284154, the best-fit plasma temperature is $kT = 0.76 \pm 0.04$\,keV with an emission measure of $EM = 9.79^{+0.51}_{-0.48} \times 10^{52}$\,cm$^{-3}$. The resulting unabsorbed X-ray luminosities are $L_{\rm X} (0.3\text{--}10\,\text{keV}) = 7.59^{+0.39}_{-0.35} \times 10^{29}$\,erg\,s$^{-1}$ and $L_{\rm X} (0.1\text{--}2.4\,\text{keV}) = 9.75^{0.54}_{-0.52}  \times 10^{29}$\,erg\,s$^{-1}$. 

For ATO\,J061.3032+20.1301, the fit yields $kT=0.85\pm0.04$\,keV, an emission measure of $EM = 1.01 \pm 0.05 \times 10^{53}$\,cm$^{-3}$, corresponding to unabsorbed luminosities of $L_{\rm X} (0.3\text{--}10\,\text{keV}) = 8.02^{+0.19}_{-0.39} \times 10^{29}$\,erg\,s$^{-1}$ and $L_{\rm X} (0.1\text{--}2.4\,\text{keV}) = 1.03 \pm 0.05 \times 10^{30}$\,erg\,s$^{-1}$.

\section{ASAS-SN light curve interpolation through Gaussian process regression} \label{sec:gpasas}

We modelled the ASAS-SN light curve collected during the timespan covered by the Swift and JVLA campaigns through a GP regression, in order to exploit the high GP capability to make prediction of the optical flux values at the epochs of the X-ray/NUV and radio observations, for which there are no simultaneous ASAS-SN data. We adopted the quasi-periodic covariance matrix \citep[e.g.][]{haywood2014MNRAS.443.2517H} with the generic element defined as  
\begin{gather} 
\label{eq:eqgpqpkernel}
k_{QP}(t, t^{\prime}) = A^2\exp\Bigg[-\frac{(t-t^{\prime})^2}{2\tau^2} - \frac{\sin^{2}(\pi(t-t^{\prime})/P)}{2w^2}\Bigg] +  \\
+\, \left(\sigma^{2}_{\rm ASAS-SN}(t)+\sigma^{2}_{\rm jit}\right)\delta_{t, t^{\prime}}\; \nonumber
\end{gather}
as the optimal GP kernel to model the periodic variability seen in the optical data, modulated over the stellar rotation period. In Eq. (\ref{eq:eqgpqpkernel}), $t$ and $t^{\prime}$ represent two different epochs of observations, $\sigma_{\rm ASAS-SN}$(t) is the uncertainty in the \textit{g}-band flux at epoch $t$, and $\delta_{t, t^{\prime}}$ is the Kronecker delta. Uncorrelated noise, either instrumental and/or astrophysical, is taken into account by adding a constant jitter term $\sigma_{\rm jit}$ in quadrature to $\sigma_{\rm ASAS-SN}$(t). The GP hyper-parameters are $A$, which denotes the scale amplitude of the correlated signal; $P$ is the stellar rotation period, which is well constrained in the case of V1298\,Tau; $w$, which describes the ``weight'' of the rotation period harmonic content within a complete stellar rotation (i.e. a low value of $w$ indicates that the periodic variations contain a significant contribution from the harmonics of the rotation period); and $\tau$, which represents the decay timescale of the correlations and is related to the temporal evolution of the magnetically active regions responsible for the correlated signal observed in the light curve. 

\begin{figure}
    \centering
    \includegraphics[width=0.5\textwidth]{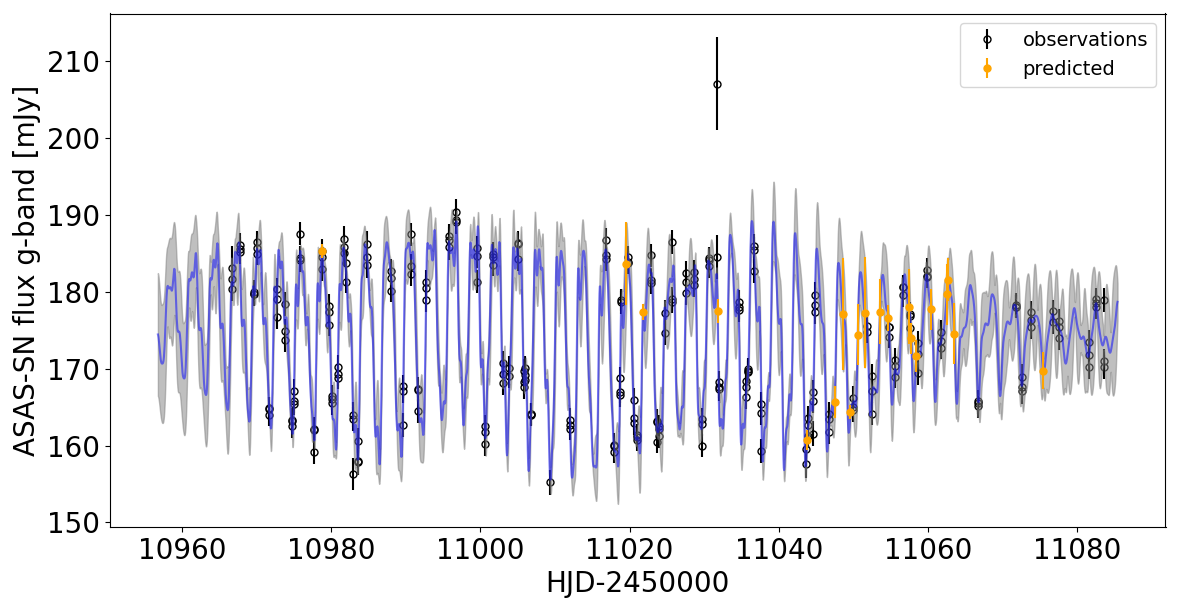}
    \caption{ASAS-SN optical light curve and predicted values at the epochs of JVLA observations. The ASAS-SN optical light curve (black dots) includes the time span of the radio campaign. The best-fit model based on a GP quasi-periodic regression is shown by a blue line (with the $\pm1\sigma$ confidence interval indicated by a grey area), and it was used to predict the optical fluxes at the epochs of JVLA observations (orange dots).}
    \label{fig:gp_light_curve}
\end{figure}

We used the publicly available \texttt{python} module \texttt{george} v0.2.1 \citep{Ambikasaran2015} to perform the GP 1-dim regression. We used the publicly available Monte Carlo (MC) nested sampler and Bayesian inference tool \texttt{MultiNest v3.10} (e.g., \citealt{Feroz2019}), through the \texttt{pyMultiNest} wrapper \citep{Buchner2014}, to derive the posterior distributions for the free (hyper-)parameters of the model, for which we adopted the following uninformative priors: $\mathcal{U}$(0,1000) mJy for $\sigma_{\rm jit}$; $\mathcal{U}$(-1000,1000) mJy for the offset in flux; $\mathcal{U}$(0,1000) mJy for the GP amplitude $A$;  $\mathcal{U}$ (2.7,3.2) day for $P$; $\mathcal{U}$(0,2) for $w$; $\mathcal{U}$(0,10) for $\tau$ in the natural logarithm space. The best-fit GP model to JVLA data is shown in Fig. \ref{fig:gp_light_curve}. 

\section{Starspot modeling}
\label{sec:spot_modeling}

We analysed the ASAS-SN photometric time series by applying the \texttt{PAStar} model \citep{Petralia2025} to the best-fit model based on a GP quasi-periodic regression. The model describes stellar surface inhomogeneities, specifically in the form of spots and faculae, to account for the observed variability in photometric and spectroscopic time series. To determine the best-fit parameters and their associated uncertainties, \texttt{PAStar} employs the \texttt{MultiNest} Bayesian framework \citep{Feroz2009}, through the \texttt{pyMultiNest} wrapper \citep{Buchner2014}. The fitted parameters include the initial positions of the features at the start of the time series (latitudes and longitudes) and their respective radii.

Several stellar parameters were kept fixed during the analysis, including the rotational period and the spot/facula contrast ratios relative to the photosphere. For V1298~Tau, we assumed that the variability is dominated by spots and consequently neglected the contribution of faculae. To ensure sufficient spatial resolution, the surface element size was set to $0.45^\circ$, which corresponds to spanning the longitude range ($\Delta$$\phi$=[$0$--$2\pi$]) with $800$ pixels, with the same element size applied to latitudes ($\Delta$$\theta$=[$0$--$\pi$]). Assuming a stellar radius of $1.24~R_\odot$ \citep{Magrini2022,Tsantaki2025}, this grid size corresponds to a physical pixel size of $\sim 7000$~km at the centre of the stellar disk, which ensures an accurate description of the large filling factors expected for active stars \citep{Berdyugina2005}. A comprehensive list of the assumed stellar parameters is provided in Table~\ref{tab:table_spotmodeling}.

To evaluate the spot contrast ratio ($\Delta F_{\text{spot}}$), we integrated \texttt{PHOENIX} synthetic spectra \citep{Claret2012,Claret2013} within the Python package \texttt{ExoTethys} \citep{Morello2020} for both the photosphere and the spots across the Sloan \textit{g}-band, yielding a value of $\sim 0.166$. Four terms limb-darkening coefficients \citep{Claret2000} were employed and computed within the same framework \texttt{ExoTethys}. For these calculations, solar metallicity was assumed due to the current limitations of the \texttt{ExoTethys} package in evaluating limb-darkening coefficients for non-solar metallicities. The Bayesian framework was configured to explore the parameter space, defined by the priors listed in Table~\ref{tab:priors_spotmod}, using $2000$ live points with a sampling efficiency of $0.5$.

Currently, the code does not account for the intrinsic evolution of active regions (e.g., variations in spot radii). Therefore, to ensure a self-consistent analysis, we divided the dataset into independent segments, each spanning the duration of a single stellar rotation. Specifically, each analysis window was sequentially shifted by half a rotation period to guarantee a temporal overlap and verify the consistency of the resulting solutions. This approach yielded a total of 68 independent fits across the ASAS-SN time series, covering the exact time period during which the radio observations listed in Table~\ref{tab:radioresults} were obtained.

The number of spots required by the model to describe the observed variability is determined by running the analysis sequentially with an increasing number of spots and subsequently comparing their Bayesian log-evidences ($Z$). If the more complex model (i.e., with a higher number of spots) yields a higher log-evidence by at least 5 ($\Delta \log Z \ge 5$), we continue to increase the complexity; otherwise, the process is terminated, and the current model is selected as the best fit. As a result, a three-spot model was found to be sufficient to explain the GP best-fit in all analysed windows.

\begin{figure*}
    \centering
    \includegraphics[width=0.95\textwidth]{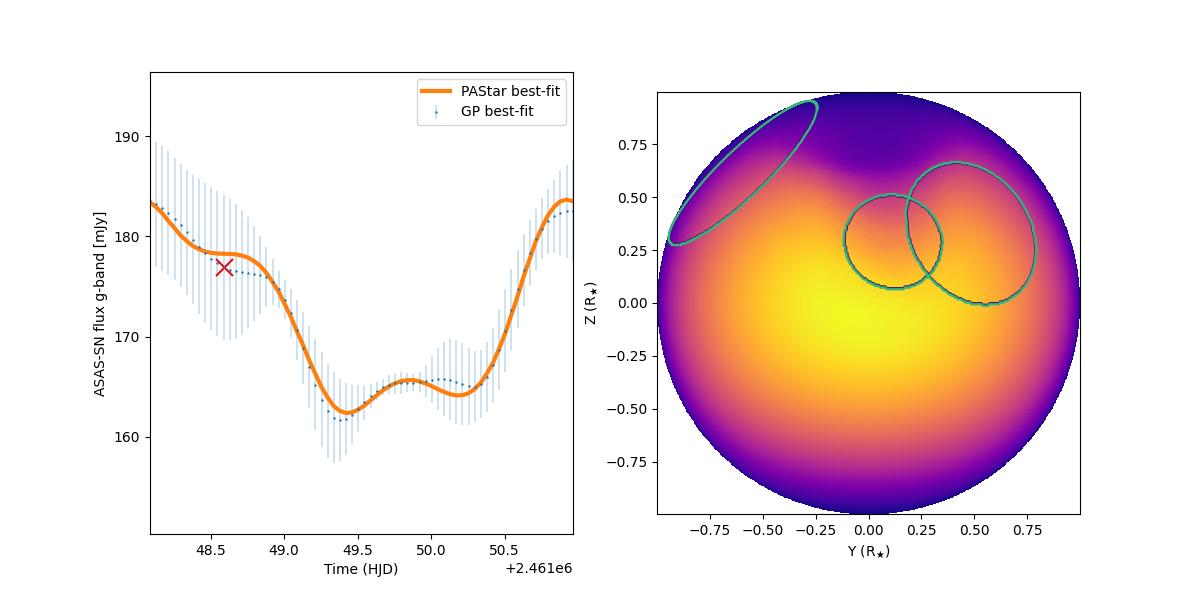 }
    \caption{ASAS-SN optical light curve and best-fit \texttt{PAStar} models. 
\textit{Left panel:} The points and error bars represent the best-fit solution obtained from the quasi-periodic GP regression, while the solid orange line shows the \texttt{PAStar} model solution for the analysed rotation period. \textit{Right panel:} Synthetic stellar surface produced by the \texttt{PAStar} model at the epoch indicated by the red cross marker in the left panel, corresponding to the simulated time closest to the brightest radio detection, observation 7 (see Table \ref{tab:radioresults}). For visualization purposes, the background yellow-to-purple gradient of the stellar disk represents the projected area of the surface elements, rather than the physical limb-darkening effect. The darker, diffuse regions are generated by superimposing one thousand spot configurations randomly sampled from the posterior distribution, illustrating the probability density and uncertainties of the spots. Finally, the open cyan circles delineate the exact locations and sizes of the maximum a-posteriori spot solutions.}
    \label{fig:result_spot_model}
\end{figure*}

Fig.~\ref{fig:result_spot_model} shows an example of the model outputs, comparing the GP best-fit light curve with the solutions obtained from \texttt{PAStar}. From the complete set of analysed observational windows, we derived the time-varying projected filling factors; in the case of overlapping windows, we adopted the values from the subsequent period.

\begin{figure}
    \centering
    \includegraphics[width=0.5\textwidth]{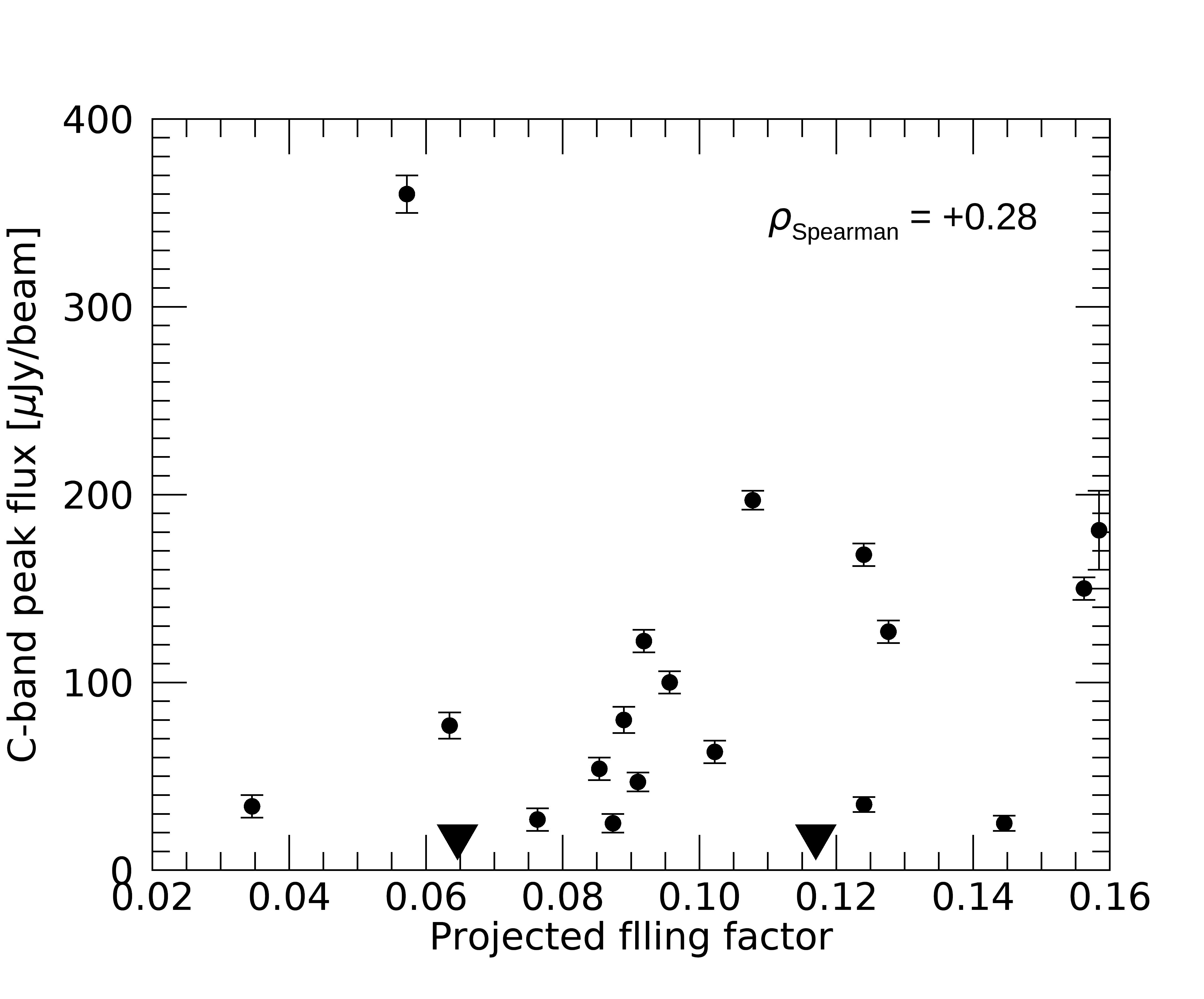}
    \caption{Correlation between the radio fluxes and the projected filling factors inferred from  \texttt{PAStar} at the corresponding epochs of radio observations. Triangles correspond to the radio peak flux $3\sigma$ upper limits, which were excluded from the calculation of the Spearman's rank correlation coefficient.}
    \label{fig:radiovsff}
\end{figure}

To investigate the connection between stellar surface inhomogeneities and chromospheric/coronal activity, we analysed the relation between the C-band radio peak flux and the projected filling factors, evaluated at the simulation epochs closest to each radio observation. Fig.~\ref{fig:radiovsff} displays the corresponding scatter plot. We find a weak positive linear correlation between the two quantities, after removing the two radio flux upper limits, quantified by a Spearman's rank correlation coefficient of $\rho_{\rm Spearman} = +0.28$. We note that the correlation turns to moderate ($\rho_{\rm Spearman}=+0.44$) after removing the radio-loudest data point. If treating this point as an outlier (see discussion in Sect. \ref{sec:corr_radio_optical}), our simulations may indicate that the observed radio emission is moderately coupled with the star's spot coverage. The tentative but not sound correlation is compatible with the radio-optical anti-correlations shown in Sect. \ref{sec:corr_radio_optical}.

\begin{table}
\centering
\small
\caption{Fixed parameters to set the {\tt PAStar} model of V1298\,Tau. }
\label{tab:table_spotmodeling}
\begin{tabular}{c c}        
\hline\hline                 
Parameter & Value \\
\hline

T$_{\bigstar}$(K)\tablefootmark{a}& 5100 \\
$\Delta$T$_{spot}$ (K) & 1000  \\
log g (dex)\tablefootmark{a} & 4.18 \\
$\Delta$F$_{spot}$ & 0.166 \\
$P$ (d)\tablefootmark{b} & 2.89 \\
$\Delta$$\phi$,$\Delta$$\theta$ (deg) & 0.45 \\
\hline
Limb darkening Coefficients\tablefootmark{c} & \\
\hline

Spot & 0.49408133, -0.14557458, \\
& 0.67515952, -0.101343 \\
Photosphere & 0.7552619, -1.08272447, \\
& 1.97910774, -0.7261707 \\

\hline
\hline
\end{tabular}
\tablefoot{\tiny
\tablefoottext{a}{\cite{Magrini2022,Tsantaki2025}},
\tablefoottext{b}{This work},
\tablefoottext{c}{\cite{Morello2020,Claret2000}}
}

\end{table}

\begin{table}
\centering
\small
\caption{Prior distributions to set the {\tt PAStar} model of V1298\,Tau.}
\label{tab:priors_spotmod}
\begin{tabular}{c c}        
\hline\hline                 
Parameter & Value \\
\hline
Flux scale (e/s) & U($F_{max}-\sigma$,$F_{max}+\sigma)$ \\
$\sigma_{\rm jit}$ & $\mathcal{LU}$($j_*/100,j_*$)  \\
$\Delta \phi$ & $\mathcal{U}$(0,2$\pi$)\\
$\Delta \theta$ & $\mathcal{U}$(0,$\pi$/2)\\
\hline
\hline
\end{tabular}
\tablefoot{\tiny{$F_{\max}$ represents the maximum value of the GP fit within the time series, and $\sigma$ denotes its associated uncertainty. The parameter $j_*$ is computed independently for each fit and is defined as three times the median of the errors exceeding the 80$^{\rm th}$ percentile threshold, providing a metric to scale the jitter relative to the observational errors. }
}
\end{table}

\end{document}